\documentclass[reprint,aps,prd,amsmath,amssymb,superscriptaddress,showpacs,floatfix]{revtex4-1}
\usepackage{graphicx}
\usepackage[hidelinks]{hyperref}

\newcommand{\cena}{Centaurus~A}

\newcommand{\degree}{\ensuremath{^{\circ}}}
\newcommand{\arcmin}{\ensuremath{^{\prime}}}

\newcommand{\eqn}{Eq.}

\newcommand{\fig}{Fig.}
\newcommand{\figs}{Figs.}
\newcommand{\tab}{Table}

\newcommand{\sect}{Sec.}

\newcommand{\Fig}{Fig.}

\newcommand{\Sect}{Sec.}

\newcommand{\refii}[2]{\ref{#1} and~\ref{#2}}

\newcommand{\eqnref}[1]{\eqn~\ref{#1}}

\newcommand{\figref}[1]{\fig~\ref{#1}}
\newcommand{\Figref}[1]{\Fig~\ref{#1}}
\newcommand{\figrefii}[2]{\figs\ \refii{#1}{#2}}

\newcommand{\tabref}[1]{\tab~\ref{#1}}

\newcommand{\secref}[1]{\sect~\ref{#1}}
\newcommand{\Secref}[1]{\Sect~\ref{#1}}

\newcommand{\gcitep}{~\citep}

\begin{document}

% Journal abbreviations as defined in style guide for Physical Review.  Using providecommand to avoid conflicts with revtex4-1.
\providecommand{\aap}{Astron.\ Astrophys.}                  % Astronomy and Astrophysics
\providecommand{\aj}{Astron.\ J.}                     % Astronomical Journal
\providecommand{\ajph}{Aust.\ J.\ Phys.} % Australian Journal of Physics
\providecommand{\anchem}{Anal.\ Chem.}   % Analytical Chemistry
\providecommand{\ao}{Appl.\ Opt.}         % Applied Optics
\providecommand{\apj}{Astrophys.\ J.}                   % Astrophysical Journal
\providecommand{\azh}{Astron.\ Zh.}                   % Astronomiceskij Zhurnal
\providecommand{\baas}{Bull.\ Am.\ Astron.\ Soc.}                 % Bulletin of the American Astronomical Society
\providecommand{\bell}{Bell Syst.\ Tech.\ J.} % Bell Systems Technical Journal
\providecommand{\cpc}{Comput.\ Phys.\ Commun.} % Computer Physics Communications
\providecommand{\dans}{Dokl.\ Akad.\ Nauk SSSR}  % Doklady Akademii Nauk SSSR (some old Russian journal)
\providecommand{\epsl}{Earth Planet.\ Sci.\ Lett.} % Earth and Planetary Science Letters
\providecommand{\gca}{Geochim.\ Cosmochim.\ Acta} % Cheochimica et Cosmochimica Acta
\providecommand{\ieeetit}{IEEE Trans.\ Inf.\ Theory} % IEEE Transactions on Information Theory
\providecommand{\ieeemtt}{IEEE Trans.\ Microwave Theory Tech.} % IEEE Transactions on Microwave Theory and Techniques
\providecommand{\jcomph}{J.\ Comput.\ Phys.} % Journal of Computational Physics
\providecommand{\jcp}{J.\ Chem.\ Phys.}      % Journal of Chemical Physics
\providecommand{\jgr}{J.\ Geophys.\ Res.}    % Journal of Geophysical Research
\providecommand{\jhep}{J.\ High Energy Phys.}                 % Journal of High Energy Physics
\providecommand{\mnras}{Mon.\ Not.\ R.\ Astron.\ Soc.}               % Monthly Notices of the RAS
\providecommand{\nat}{Nature}                % Nature
\providecommand{\nima}{Nucl.\ Instrum.\ Methods A} % Nuclear Instruments & Methods in Physics Research Section A
\providecommand{\pasp}{Publ.\ Astron.\ Soc.\ Pac.}                 % Publications of the Astronomical Society of the Pacific
\providecommand{\phr}{Phys.\ Rev.}           % Physical Review
\providecommand{\pra}{Phys.\ Rev.\ A}        % Physical Review A: Atomic, Molecular, and Optical Physics
\providecommand{\prb}{Phys.\ Rev.\ B}        % Physical Review B: Solid State
\providecommand{\prc}{Phys.\ Rev.\ C}        % Physical Review C: Nuclear Physics
\providecommand{\prd}{Phys.\ Rev.\ D}        % Physical Review D: Particles & Fields
\providecommand{\pre}{Phys.\ Rev.\ E}        % Physical Review E: Statistical Physics, Plasmas, Fluids, and Related Interdisciplinary Topics
\providecommand{\prl}{Phys.\ Rev.\ Lett.}    % Physical Review Letters
\providecommand{\planss}{Planet.\ Space Sci.} % Planetary and Space Science
\providecommand{\radsci}{Radio Sci.}         % Radio Science
\providecommand{\rpph}{Rep.\ Prog.\ Phys.}   % Reports on Progress in Physics
\providecommand{\rsla}{Phil.\ Trans.\ R.\ Soc.\ A} % Philosophical Transactions of the Royal Society A: Mathematical, Physical and Engineering Sciences
\providecommand{\sal}{Sov.\ Astron.\ Lett.}  % Soviet Astronomy Letters
\providecommand{\spjetp}{Sov.\ Phys.\ JETP}  % Soviet Journal of Experimental and Theoretical Physics
\providecommand{\spu}{Sov.\ Phys.\ Usp.}  % Soviet Physics Uspekhi
\providecommand{\sci}{Science}               % Science
\providecommand{\solph}{Sol.\ Phys.}         % Solar Physics
\providecommand{\zap}{Z.\ Astrophys.}        % Zeitschrift fuer Astrophysik

% Further abbreviations defined to be consistent with the above style.
\newcommand{\spjetpl}{Sov.\ Phys.\ JETP Lett.} % Sovient Journal of Experimental and Theoretical Physics Letters
\newcommand{\pla}{Phys.\ Lett.\ A}       % Physics Letters A
\newcommand{\plb}{Phys.\ Lett.\ B}       % Physics Letters B
\newcommand{\app}{Astropart.\ Phys.}     % Astroparticle Physics
\newcommand{\nar}{New Astron.\ Rev.}     % New Astronomy Reviews
\newcommand{\pasa}{Publ.\ Astron.\ Soc.\ Aust.}                 % Publications of the Astronomical Society of Australia
\newcommand{\expa}{Exp.\ Astron.}        % Experimental Astronomy
\newcommand{\privcom}{priv.\ comm.}      % private communication
\newcommand{\apjl}{Astrophys.\ J.\ Lett.}                  % Astrophysical Journal Letters
\newcommand{\jcap}{J.\ Cosmology Astropart.\ Phys.} % Journal of Cosmology and Astroparticle Physics
\newcommand{\njp}{New J.\ Phys.}         % New Journal of Physics
\newcommand{\ijmpd}{Int.\ J.\ Mod.\ Phys.\ D} % International Journal of Modern Physics D
\newcommand{\araa}{Annu.\ Rev.\ Astron.\ Astrophys.}               % Annual Review of Astronomy and Astrophysics
\newcommand{\aipcs}{AIP Conf.\ Series}   % American Institute of Physics Conference Series

\title{A limit on the ultra-high-energy neutrino flux from lunar observations with the Parkes radio telescope}

\newcommand{\adelaideuni}{School of Chemistry \& Physics, Univ.\ of Adelaide, SA 5005, Australia}
\newcommand{\atnf}{CSIRO Astronomy \& Space Science, Epping, NSW 1710, Australia}
\newcommand{\sotonuni}{School of Physics \& Astronomy, Univ.\ of Southampton, SO17 1BJ, United Kingdom}
\newcommand{\erlangenuni}{ECAP, Univ.\ of Erlangen-Nuremberg, 91058 Erlangen, Germany}
\newcommand{\astron}{ASTRON, 7990 AA Dwingeloo, The Netherlands}

\author{J.D.\ Bray}
 \affiliation{\adelaideuni}
 \affiliation{\atnf}
 \affiliation{\sotonuni}

\author{R.D.\ Ekers}
 \affiliation{\atnf}

\author{P.\ Roberts}
 \affiliation{\atnf}

\author{J.E.\ Reynolds}
 \affiliation{\atnf}

\author{C.W.\ James}
 \affiliation{\erlangenuni}

\author{C.J.\ Phillips}
 \affiliation{\atnf}

\author{R.J.\ Protheroe}
 \affiliation{\adelaideuni}

\author{R.A.\ McFadden}
 \affiliation{\astron}

\author{M.G.\ Aartsen}
 \affiliation{\adelaideuni}

\date{\today}

\begin{abstract}
 We report a limit on the ultra-high-energy neutrino flux based on a non-detection of radio pulses from neutrino-initiated particle cascades in the Moon, in observations with the Parkes radio telescope undertaken as part of the LUNASKA project.  Due to the improved sensitivity of these observations, which had an effective duration of 127~hours and a frequency range of 1.2--1.5~GHz, this limit extends to lower neutrino energies than those from previous lunar radio experiments, with a detection threshold below 10$^{20}$~eV.  The calculation of our limit allows for the possibility of lunar-origin pulses being misidentified as local radio interference, and includes the effect of small-scale lunar surface roughness.  The targeting strategy of the observations also allows us to place a directional limit on the neutrino flux from the nearby radio galaxy \cena.
\end{abstract}


\maketitle

\section{Introduction}

The spectrum of cosmic rays (CRs) extends into the ultra-high-energy (UHE; \mbox{$> 10^{18}$}~eV) regime, in which their origin is currently unknown.  Above a threshold energy of ${\sim 6} \times 10^{19}$~eV (for CR protons) they lose energy through interactions with the cosmic microwave background\gcitep{greisen1966,zatsepin1966}, known as the Greisen-Zatsepin-Kuzmin (GZK) effect.  This results in a dramatic steepening in the CR spectrum above this energy\gcitep{abraham2010}.  The few CRs observed above the GZK threshold energy must originate within the local universe, within a few times the energy-loss distance (tens of Mpc).

CRs across a range of energies are known to be produced in regions of magnetic turbulence such as supernova remnants\gcitep{ackermann2013}, where charged particles can be accelerated through interactions with magnetic fields.  Because this process starts with a low-energy particle and increases its energy to produce a CR, models of this type are described as \textit{bottom-up} models, with the details depending on the particular acceleration mechanism and the environment of the source object.  In the UHE regime, it is proposed that CRs could also originate via a \textit{top-down} mechanism in which UHECRs are decay products from hypothetical superheavy particles, with the details of the models depending on the assumed properties of these particles.

%UHECR origin models are divided into two classes: \textit{bottom-up} models, in which low-energy particles are accelerated until they reach ultra-high energies; and \textit{top-down} models, in which UHECRs are decay products from hypothetical superheavy particles.

Different UHECR origin models predict a range of different UHECR source spectra; however, the observed spectrum is modified by energy-dependent GZK attenuation and photo-disintegration of CR nuclei, making discrimination difficult.  Different models also predict different astronomical objects from which UHECRs should originate, but charged UHECRs are deflected in cosmic magnetic fields, obscuring their directions of origin.  These deflections are less significant for more energetic UHECRs, and there is a correlation between UHECR arrival directions and the positions of nearby active galactic nuclei (AGN), a commonly-proposed class of source; but these are in turn correlated with the distribution of matter in the local universe, and thus with sources for other models\gcitep{abraham2008}.  There is also a tentative correlation of UHECR arrival directions on large scales (\mbox{$\sim 20$}\degree) with the nearby AGN \cena\gcitep{abreu2010}.

Another approach for exploring the origin of UHECRs is to search for counterpart neutrinos at similar energies.  Both bottom-up and top-down origin models predict an associated UHE neutrino flux.  In addition, independent of the origin model, UHECRs undergoing GZK interactions produce pions which decay to produce additional UHE neutrinos.  Since neutrinos do not interact significantly with intervening matter or radiation fields, these neutrinos retain information about the source UHECR spectrum\gcitep{protheroe2004}.  In addition, since they are uncharged, they are not deflected by magnetic fields, and their arrival directions correspond directly to their directions of origin.  Consequently, the detection of even a single UHE neutrino (with sufficient angular resolution) could identify a source of UHECRs, and sufficient detections to measure the UHE neutrino spectrum would provide a strong test of competing UHECR origin models.

Efforts to detect UHE neutrinos focus on searching for nanosecond-scale radio pulses produced when they interact in dense media: the interaction initiates a particle cascade which develops an excess negative charge primarily through entrainment of electrons from the medium, resulting in coherent radiation at wavelengths larger than the width of the cascade ($\lambda \gtrsim 10$~cm; i.e.\ radio), directed at the Cherenkov angle.  This burst of coherent radiation was predicted by \citet{askaryan1962}, and we refer to it as an Askaryan pulse.  The effect requires that the particle cascade occur in a radio-transparent medium; it has been observed in laboratory experiments with target media of silica sand\gcitep{saltzberg2001}, rock salt\gcitep{gorham2005} and ice\gcitep{gorham2007}.

UHE neutrino radio-detection experiments have been conducted with terrestrial ice as a target medium, monitored with radio antennas embedded within the ice \citep[RICE]{kravchenko2006}, suspended from a high-altitude balloon \citep[ANITA]{gorham2009b}, or mounted on a satellite \citep[FORTE]{lehtinen2004}.  They have also been conducted with the target medium being the lunar regolith (with properties similar to silica sand), monitored with ground-based radio telescopes as suggested by \citet{dagkesamanskii1989}.  Due to the size of the Moon, this latter technique offers the largest potential experimental aperture; however, it requires the Askaryan pulse to be bright enough to be visible at the distance of the Moon ($\sim 3.8 \times 10^8$~m), which makes it sensitive only to the most energetic neutrinos.  This makes it well-suited to testing top-down UHECR origin models, as these typically predict neutrino spectra extending to higher energies than those from bottom-up models or GZK interactions \citep[e.g.][]{protheroe1996b,berezinsky2011,lunardini2012}.  In addition to neutrinos, this technique has some capacity to detect UHECRs directly\gcitep{terveen2010,jeong2012}, but in this article we consider the sensitivity to neutrinos only.

The LUNASKA (Lunar UHE Neutrino Astrophysics with the Square Kilometre Array) project aims to develop this technique for future use with the Square Kilometre Array\gcitep{carilli2004}, a radio telescope with a planned sensitivity exceeding all present instruments.  As part of this project, we have conducted an experiment employing this technique with the Parkes radio telescope, searching for lunar-origin Askaryan pulses as a tracer of UHE particle cascades in the lunar regolith.  \Secref{sec:design} contains an overview of the design of our experiment, and \secref{sec:obs} summarizes our observations and analysis; full details of our experimental procedure are published in a separate article\gcitep{bray2014a}.  As in previous experiments of this type, we do not detect any lunar-origin radio pulses, which establishes a limit on the UHE neutrino flux, through the procedure described in \secref{sec:sims}.  Due to our observational strategy, we are able to place limits both on the diffuse flux (\secref{sec:diffuse}) and on the directional flux from \cena\ (\secref{sec:directional}).  Finally, \secref{sec:conclusion} summarizes our major results.

\section{Experiment design}
\label{sec:design}

\begin{figure}
 \centering
 \includegraphics[width=\linewidth]{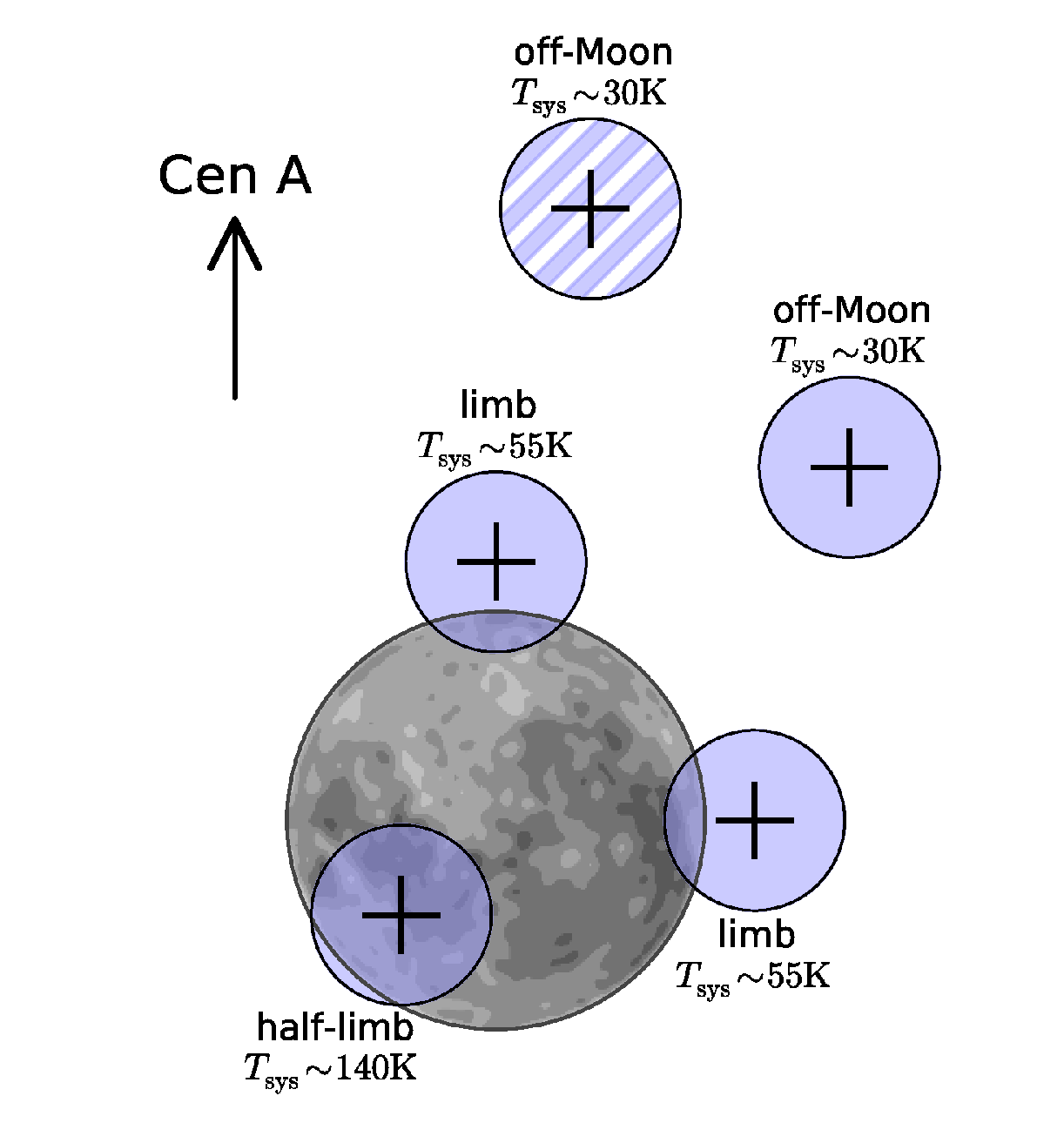}
 \caption{Typical pointings (solid shading) of telescope beams relative to the Moon in our observations, with approximate system temperatures $T_{\rm sys}$ for each beam.  The limb beams are more sensitive than the half-limb beam; the off-Moon beam is even more sensitive, but useful only for identifying RFI.  Crosses in each beam indicate the orientation of the orthogonal linear polarizations.  The receiver orientation was maintained such that \cena\ was always in the direction shown.  For some of the observing time, the half-limb beam was replaced with an additional off-Moon beam (striped).}
 \label{fig:config}
\end{figure}

Our experiment was conducted the Parkes radio telescope, a single 64~m parabolic antenna, using its 21~cm multibeam receiver\gcitep{staveley-smith1996}, with a radio frequency range of 1.2--1.5~GHz.  This receiver is capable of operating thirteen simultaneous beams on the sky, of which we used four beams at any one time, as shown in \figref{fig:config}.  Two or three of these beams were placed near the limb of the Moon, matching the expected position of a detectable Askaryan pulse\gcitep{james2009f}, while the remaining one or two beams were directed away from the Moon and used to identify spurious signals from radio frequency interference (RFI).  One limb beam was always maintained at the point on the limb closest to \cena\ in order to maximize the sensitivity to neutrinos (or cosmic rays) from that direction, as done by \citet{james2010}.

Real-time processing of the signal, after downconversion to an intermediate frequency range of 50--350~MHz, was carried out with the Bedlam backend\gcitep{bray2012}.  A buffer of 4~$\mu$s of sampled baseband data was maintained for each of the eight input channels, corresponding to two polarizations on each of four beams.  If a peak exceeding a trigger threshold was detected in any on-Moon beam, and no peak was simultaneously detected on any other beam, the contents of these buffers were copied to permanent storage.  This latter anticoincidence condition excluded the majority of RFI pulses, which typically entered from sidelobes of the antenna, and so were detected on multiple beams.

\subsection{Signal optimization}
\label{sec:sigproc}

The amplitude of a lunar-origin Askaryan pulse is decreased by dispersion as it passes through the ionosphere.  The degree of this dispersion is determined by the ionospheric total electron content (TEC); specifically, by the column density of plasma along the line of sight to the Moon, called the slant TEC (STEC) and typically measured in TEC units (TECU).  The Bedlam backend implements a digital dedispersion filter which compensates for this, but was limited by our imperfect knowledge of the STEC.  The maximum deviation between the filter setting, based on real-time ionosonde measurements, and the more precise STEC measurements available retrospectively from Global Positioning System (GPS) data, was 9.0 TECU, which determines the real-time loss of signal strength from dispersion.  The loss of signal strength in full retrospective processing is determined by the maximum uncertainty in the GPS-derived STEC, which was 3.8 TECU.  The maximum value of the STEC at any point during our observations was 23.5 TECU.

Digital sampling at a finite rate can reduce the amplitude of an analog pulse, as the sampled points will not generally coincide with the peak of the pulse\gcitep{james2010}.  This effect can be counteracted by interpolating between the sampled points.  The Bedlam backend implements two-fold interpolation, reconstructing one point between every pair of sampled points, effectively doubling the real-time sampling rate.  In retrospective processing, we performed 32-fold interpolation, almost completely negating this effect.

The phase of a pulse is randomized by the process of frequency downconversion, with a consequent potential loss of amplitude\gcitep{bray2012}.  In real time, no effort was made to compensate for this effect.  In retrospective processing, we formed the signal envelope, which restores the full amplitude of the pulse for any possible phase, at the cost of increasing the noise level.

The individual contributions of the effects described above to the signal loss are shown in \tabref{tab:sigloss}, for both real-time processing and retrospective processing, compared to the case if no processing had been applied.  The real-time processing was sufficient to cause any significant event to exceed the trigger threshold, allowing it to be recovered for later analysis.  The retrospective processing applied to these stored events was sufficient to reconstruct almost the maximum pulse amplitude, with a maximum signal loss of only 0.4\%.

\begin{table}
 % Values obtained from co-output of sensloss.pdf.
 \centering
 \caption{Simulated loss of signal strength due to ionospheric dispersion, finite sampling rate and unknown pulse phase; with no processing applied, with the real-time processing used in this experiment, and with the retrospective processing applied to recorded data.  Dispersion is based on the listed STEC; see text for details.}
 \begin{ruledtabular}
  \begin{tabular}{lrrrrr}
    & \multicolumn{1}{c}{STEC} & \multicolumn{4}{c}{Maximum signal loss} \\
   \cline{3-6}
     Processing & \multicolumn{1}{c}{(TECU)} & \multicolumn{1}{c}{dispersion} & \multicolumn{1}{c}{sampling} & \multicolumn{1}{c}{phase} & \multicolumn{1}{c}{total} \\
   \hline
    none & 23.5 \hspace{6pt} & 15.0\% & 21.6\% & 17.9\% & 41.9\% \\
    real-time & 9.0 \hspace{6pt} & 2.3\% & 5.6\% & 17.9\% & 23.1\% \\
    retrospective & 3.8 \hspace{6pt} & 0.4\% & $< 0.1$\% & $< 0.1$\% & 0.4\% \\
  \end{tabular}
 \end{ruledtabular}
 \label{tab:sigloss}
\end{table}

\section{Observations and analysis}
\label{sec:obs}

Our observations were conducted in runs of 3--5 days once per lunar orbit, when the Moon was closest to \cena, with a total of six runs from April to September 2010.  After subtracting time spent calibrating or otherwise excluded, we observed the Moon for a total of 148.7 hours.  We recorded candidate events typically at a rate of 1--2~Hz, with a total of 794,568 events stored for later analysis.  The majority of these events contained only fluctuations from the thermal noise background, but the events containing high-significance peaks were dominated by RFI pulses, despite the majority of such pulses being excluded by our real-time anticoincidence filter, at a ratio of \mbox{$\sim 150$:1}.  For full details of our observations, and of the analysis described below, see Ref.\gcitep{bray2014a}

After applying the full retrospective processing described in \secref{sec:sigproc} to optimize the signal-to-noise ratio, we applied a series of cuts to remove the RFI events.  These retrospective cuts refined the real-time anticoincidence filter, applying more stringent criteria to exclude pulses detected in multiple beams, at the cost of incurring an increased probability of incorrectly excluding non-RFI events.  We also applied cuts to exclude events with strong narrow-band RFI, which could saturate the analog-to-digital converters; events with a pulse width in excess of 10~ns, which is significantly greater than the expected duration of a band-limited Askaryan pulse; and events which occurred within 10~s of other probable RFI events, as these events tended to occur in bursts on this timescale.

The effective observing time of the experiment is reduced by the possibility of a non-RFI event being incorrectly excluded by these cuts or by the real-time anticoincidence filter, and by the dead time of 51~ms for storing each detected event, during which we were unable to respond to further events.  After allowing for these losses, the effective observing time was 85.5\% of the total on-Moon time, or 127.2 hours, of which 27.8 hours lacked the half-limb beam shown in \figref{fig:config}.

%\begin{figure}
% \centering
% \includegraphics[width=\linewidth]{hist_byfilt_rough}
% \caption{Distribution of amplitudes of events (bin width 0.1$\sigma$), before and after the retrospective application of cuts to remove RFI, relative to the expected distribution from pure thermal noise.  A greater number of events, by a factor of \mbox{$\sim 150$}, were rejected by the real-time anticoincidence filter, and not recorded.  The excess of high-amplitude recorded events is removed by the cuts, with the remainder being consistent with expectations.}
% \label{fig:hist_byfilt_rough}
%\end{figure}

After all cuts, no events remained with peak amplitudes in excess of 8.6$\sigma$, which is consistent with the expected thermal noise background.
% (see \figref{fig:hist_byfilt_rough}).  
We calibrated the flux sensitivity of our experiment based on the noise power in recorded events when pointing at the centre of the Moon, using a model of the lunar thermal emission\gcitep{moffat1972}, with a precision in the voltage domain of $\pm 4.5$\% (random) $\pm 4$\% (systematic).  Based on this calibration, the peak observed amplitude is equivalent, for a pulse originating at the centre of a beam and detected in either polarization, to a threshold of 0.0047 $\mu$V/m/MHz for the limb beams, or 0.0074 $\mu$V/m/MHz for the half-limb beam.  These values improve on the most sensitive previous lunar radio experiment, which had a threshold of 0.0145--0.016 $\mu$V/m/MHz\gcitep{james2010}.

\section{Neutrino aperture calculation}
\label{sec:sims}

The absence of radio pulses with amplitudes above the thresholds given above allows us to set a limit on the UHE neutrino flux.  We use the Monte Carlo simulation of \citet{james2009b} to determine the effective aperture of this experiment to fluxes of UHE neutrinos, both diffuse and directional.  In the simulation we adopt a lunar diameter of 32.0\arcmin, its median value during our observations, and an Airy disk beam shape.  We also assume constant sensitivity across the frequency range 1.2--1.5~GHz, neglecting the minor (\mbox{$\sim 10$\%}) variation.

We implement in the simulation the anticoincidence logic used to remove RFI, which can inappropriately reject lunar-origin pulses which are sufficiently intense to be detected directly in one beam and through a sidelobe of another.  This anticoincidence rejection effect is controlled by the strictest anticoincidence cut, which excludes any pulses exceeding a 4.5$\sigma$ exclusion threshold in any beam other than the triggering beam.  This effect is highly sensitive to the sidelobe pattern of the assumed Airy disk beam shape, and the receiver used in this experiment has been deliberately designed to minimize the sidelobe power\gcitep{staveley-smith1996}.  We therefore scale the exclusion threshold by the ratio between the beam power at the first sidelobe of an Airy disk (1.7\%) and a representative value for the true sidelobe power (0.5\%) we derive from measurements of the beam pattern\gcitep{staveley-smith2009}, taking the square root of this ratio to express it in terms of electric field rather than power.  Note that a nanosecond-scale pulse detected through a sidelobe will be spread out in time, further reducing its peak amplitude relative to the value predicted by beam pattern measurements conducted with a continuous radio source; we neglect the minor benefit from this effect.

For ease of computation, we include in the simulation only the radio emission from the hadronic cascade produced directly by a neutrino-nucleon interaction, in either the charged-current or neutral-current case.  This neglects the additional emission from charged leptons produced in charged-current interactions, which makes a \mbox{$\sim 10$}\% contribution to the experimental neutrino aperture\gcitep{james2009b}, smaller than the uncertainty from other effects \citep[Apps.\ A \& B]{james2010}.  Other parameters, concerning the neutrino-nucleon cross-section \citep[from][]{gandhi1998} and the radio-transparency of the lunar megaregolith, are the same as those taken by \citet{james2010} and \citet{james2011}.

\subsection{Small-scale lunar surface roughness}
\label{sec:roughness}

Lunar surface roughness influences the sensitivity of lunar radio experiments through distinct effects at different scales.  Roughness on scales larger than the length of a particle cascade determines the local surface slope and thus how efficiently the radio emission from a cascade can escape the surface, which is taken into account by most models, including the simulations of \citet{james2009b} used here.  Roughness on scales smaller than the cascade length, but larger than the radio wavelength, causes the emission from the cascade to be scattered in multiple directions longitudinally along its axis, decreasing its intensity but increasing the solid angle into which it is beamed\gcitep{james2010}.  (Scattering also occurs laterally to the cascade axis, but this has a much reduced effect, as the emission beaming already has a large lateral extent.)  The result is a decrease in the aperture for lower-energy neutrinos, for which the intensity of the Askaryan pulse is reduced below the detection threshold; but an increase in the aperture for higher-energy neutrinos, for which the Askaryan pulse remains above the detection threshold, and has a greater chance of being detected because of its increased solid angle.  Both of these effects are more pronounced for high-frequency observations, as the wavelength is a smaller fraction of the cascade length; for an experiment for which the wavelength is similar to the cascade length, such as NuMoon\gcitep{buitink2010}, both effects should be less significant\gcitep{james2012}.

No complete method to determine the effects of small-scale surface roughness has yet been developed.  A full model would require modelling roughness over a wide range of scales, including the diffractive and decoherence effects of propagation through the surface on the field amplitude over a broad bandwidth, and doing so for a wide range of randomised surfaces.  Indeed, the only existing estimate of the effects of small-scale surface roughness on lunar Askaryan pulse detection was made in the analysis of a previous LUNASKA experiment by \citet[][App.\ B]{james2010}, whose model we apply here.  They calculate the aperture as a linear combination of two extreme cases,
 \begin{equation}
  A(E) = r A_{\rm S}(E) + (1-r) A_{\rm R}(E) \label{eqn:roughapgen}
 \end{equation}
where the parameter $r$ describes the relative contributions from $A_{\rm S}(E)$, the aperture for the case of a locally-smooth moon with no small-scale surface roughness, determined by the previous simulations of \citet{james2009b}; and $A_{\rm R}(E)$, the aperture that results when the radio emission from the cascade is considered to be refracted through $N_{\rm S}$ separate surface elements with independent slopes.  Since the surface slope will have some degree of correlation over the length of the cascade, the true aperture should lie between these two extremes, and $r$ reflects the strength of that correlation.  The effective number of surface elements $N_{\rm S}$ is taken to be the ratio between the cascade length and the radio wavelength, implying a series of surface elements along the length of a near-surface cascade; this parameter is determined primarily by the radio frequency, with only a weak, logarithmic dependence on the cascade energy.

\begin{figure}
 \centering
 \includegraphics[width=\linewidth]{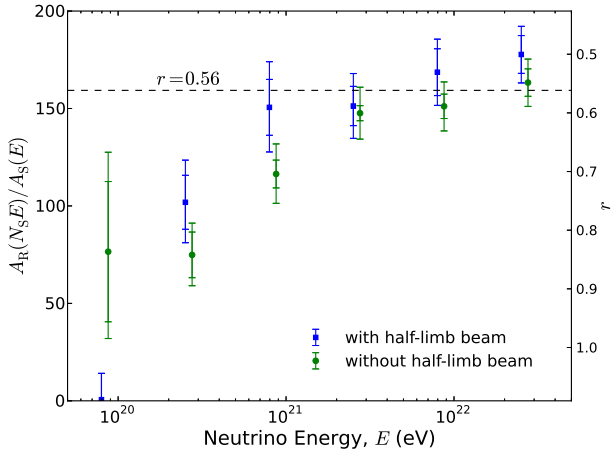}
 \caption{Simulation of the offset aperture ratio $A_{\rm R}( N_{\rm S} E ) / A_{\rm S}(E)$ in order to determine the parameter $r$, which quantifies the effects of small-scale surface roughness as described in the text (see \eqnref{eqn:roughapgen}).  Points have been slightly offset in energy for clarity.  The error bars reflect the statistical uncertainty in the simulation data (inner bars) and with the additional uncertainty from taking $N_{\rm S} = 12 \pm 1$ (outer bars).  The value \mbox{$r = 0.56$} is a fit to the range $E > 10^{21}$~eV, for which small-scale surface roughness effects are significant.}
 \label{fig:roughrat}
\end{figure}

\citet[][Eq.\ B9]{james2010} show that an appropriate value for $r$ can be determined by fitting the offset aperture ratio \mbox{$A_{\rm R}(N_{\rm S} E) / A_{\rm S}(E)$}: the amplitude of the electric field transmitted through a single surface element will be reduced by a factor $N_{\rm S}$, and the electric field is proportional to the primary particle energy $E$, hence the offset in energy by a factor $N_{\rm S}$ between the cases with and without surface elements.  We use the results of their simulations for $A_{\rm R}(E)$ to repeat their approach, as shown in \figref{fig:roughrat} (equivalent to their Fig.\ 18).  We take $N_{\rm S} = 12$, scaled from their value of $N_{\rm S} = 13$ for the small difference in the central radio frequency between the two experiments, and otherwise perform the same procedure, finding $r = 0.56 \pm 0.04$.  This differs slightly from the value of $r = 0.70 \pm 0.06$ found by \citet{james2010}, which we attribute to the significant differences between the two experiments (and hence between their representations in the simulations).

As the simulations to determine $A_{\rm R}(E)$ are computationally expensive, we make use of the further result from \citet[][Eq.\ B10]{james2010} that the aperture from \eqnref{eqn:roughapgen} can be approximated by
 \begin{equation}
  A(E) = r A_{\rm S}(E) + (1-r) k_{\rm L} N_{\rm S}^{2.363} A_{\rm S}(E / N_{\rm S})\label{eqn:roughapspec}
 \end{equation}
where the factor
 \begin{equation}
  k_{\rm L} = (1 + 0.075 \log_{10}N_{\rm S})^{-1}
 \end{equation}
corrects for the variation of the cascade length with the neutrino energy.  Apertures calculated with \eqnref{eqn:roughapspec} with \mbox{$r = 0.56$} and \mbox{$N_{\rm S} = 12$} (as above) are shown in \figref{fig:iso_ap}, and corresponding limits are shown in \figrefii{fig:iso_flux_ssr}{fig:cena_flux}.  These show the expected behaviour, with a decrease in aperture at low energies and a large increase in aperture at high energies; the abrupt transition between these two regimes is an artefact of the model.  However, since this model of the effects of small-scale surface roughness is less well-established than models of the neutrino aperture neglecting these effects, we show them in addition to, rather than instead of, the latter.  This model will generally tend to overestimate the effects of small-scale surface roughness\gcitep{james2010}, so it is likely that the true apertures and limits lie between the two scenarios shown.

\section{Diffuse neutrino sensitivity}
\label{sec:diffuse}

\begin{figure}
 \centering
 \includegraphics[width=\linewidth]{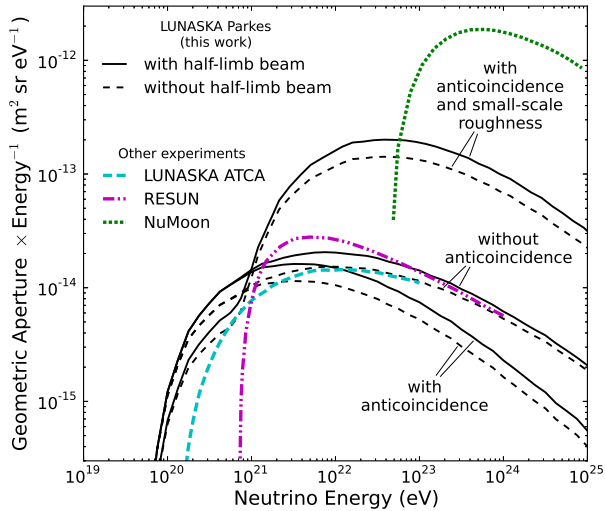}
 \caption{Geometric neutrino aperture of this experiment before inclusion of anticoincidence rejection of detected events, after allowing for this effect, and after also including the effects of small-scale lunar surface roughness (see \secref{sec:roughness}).  Apertures are also shown for other recent lunar radio experiments: a previous LUNASKA experiment with the ATCA\gcitep{james2010}, NuMoon\gcitep{buitink2010}, and RESUN\gcitep{jaeger2010}.  See text for a discussion of the various models on which these apertures are based.}
 \label{fig:iso_ap}
\end{figure}

The geometric apertures we calculate for a diffuse isotropic flux of neutrinos are shown in \figref{fig:iso_ap}.  More energetic neutrinos are more likely to produce a radio pulse sufficiently intense to be detected in multiple beams, so anticoincidence rejection reduces the effective aperture to them.  The half-limb beam improves the coverage of the Moon, but has a higher detection threshold than the limb beams, so pointing configurations that incorporate it show an increased aperture only at higher neutrino energies.

\Figref{fig:iso_ap} also shows neutrino apertures for other recent lunar radio experiments, which are based on various models.  The apertures for LUNASKA ATCA are, as for this work, based on the simulations of \citet{james2009b}.  The apertures for RESUN are from the analytic model of \citet{gayley2009}, which they show to be consistent with these simulations.  The apertures for NuMoon are from the simulations of \citet{scholten2006} as reported by \citet{buitink2010}, which are more optimistic by a factor \mbox{$\sim 10$} than those determined for the same experiment by \citet{jaeger2010} using the aforementioned analytic model; \citet{james2011} found a similar discrepancy relative to their simulations.  Nonetheless, the qualitative distinctions between these experiments --- the lower observing frequency (113--168~MHz) of NuMoon leads to a larger aperture for more energetic neutrinos, while the reduced electric field threshold for our experiment decreases the minimum detectable neutrino energy --- are predicted by all of the above models.  The major remaining uncertainties in the apertures are associated with uncertainty in the neutrino-nucleon cross-section, which may alter the aperture by a factor of \mbox{$\sim 2$}\gcitep{james2010}; and the effects of small-scale lunar surface roughness, discussed in \secref{sec:roughness}.

Note that the sensitivity calculations for these other experiments did not consider all the experimental effects taken into account for this experiment, described in \secref{sec:sigproc}.  Corrections for ionospheric dispersion were applied by LUNASKA ATCA and NuMoon, and are substantially less significant for RESUN, but the effect of the phase of an Askaryan pulse on its amplitude has been neglected by all other lunar radio experiments to date, and the signal loss due to the finite sampling rate was considered only by LUNASKA ATCA.  The loss of effective aperture due to anticoincidence rejection does not apply to LUNASKA ATCA and RESUN, which did not use this technique for RFI rejection, but does apply to the NuMoon experiment, which rejected coincident pulses between its two on-Moon beams.  Consequently, the aperture estimation for our experiment is more pessimistic than for these others, and their true sensitivity will be somewhat less than previously reported.

\begin{figure}
 \centering
 \includegraphics[width=\linewidth]{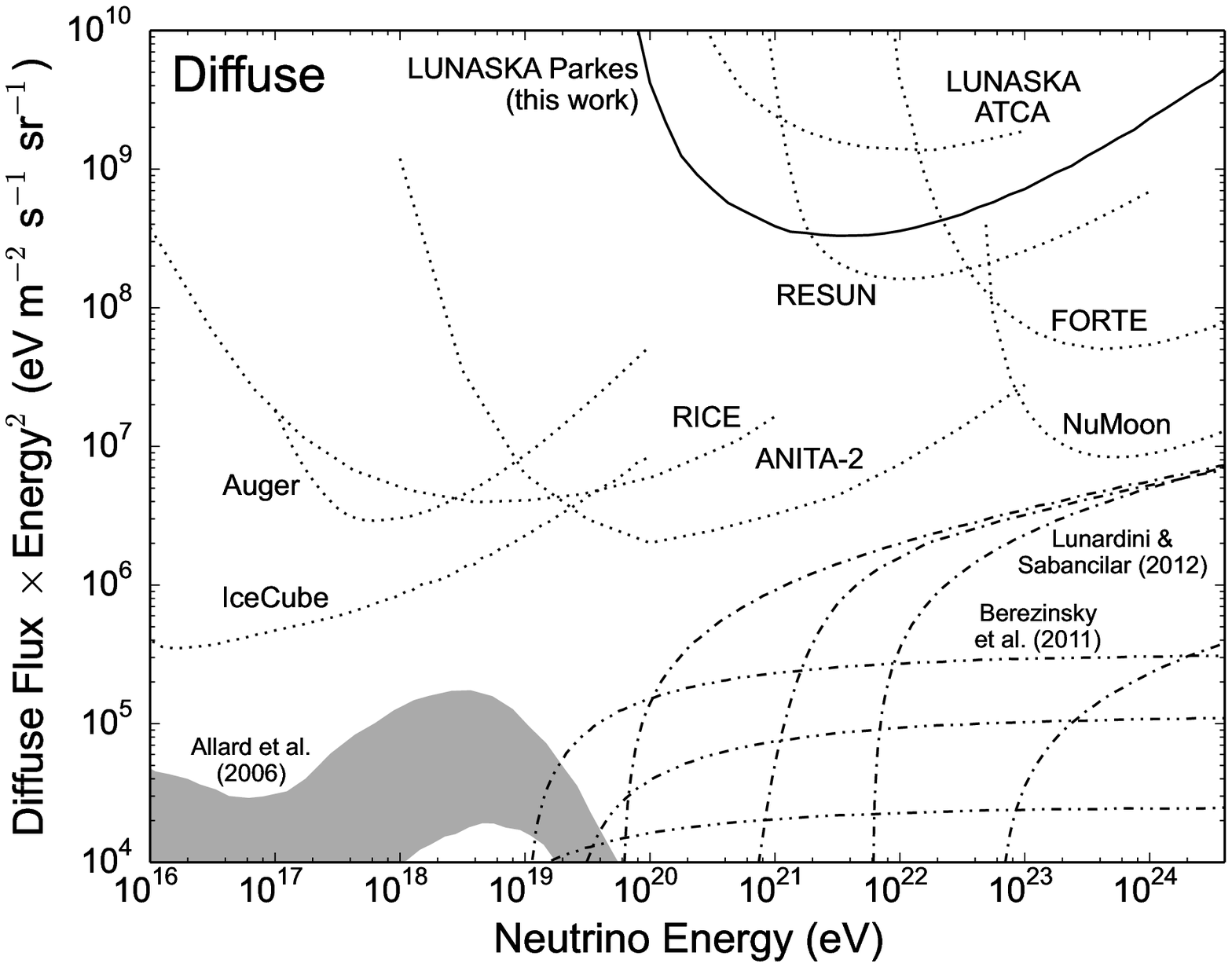}
 \caption{Model-independent limits to the diffuse neutrino flux from our experiment (solid), including the effects of anticoincidence rejection, and (dotted) for other experiments shown in \figref{fig:iso_ap}, as well as for selected non-lunar neutrino detection experiments, assuming an equal ratio of neutrino flavours: ANITA-2\gcitep{gorham2010}, RICE\gcitep{kravchenko2012}, IceCube\gcitep{aartsen2013c}, FORTE\gcitep{lehtinen2004} and the Pierre Auger Observatory, combining the exposure to both upgoing\gcitep{abreu2012} and downgoing\gcitep{abreu2011b} neutrinos.  A range of models is shown for the expected neutrino flux from GZK interactions given by \citet{allard2006} (shaded), and for top-down UHECR origin models from \citet{lunardini2012} (dash-dotted) and \citet{berezinsky2011} (dot-dash-dotted).}
 \label{fig:iso_flux}
\end{figure}

\begin{figure}
 \centering
 \includegraphics[width=\linewidth]{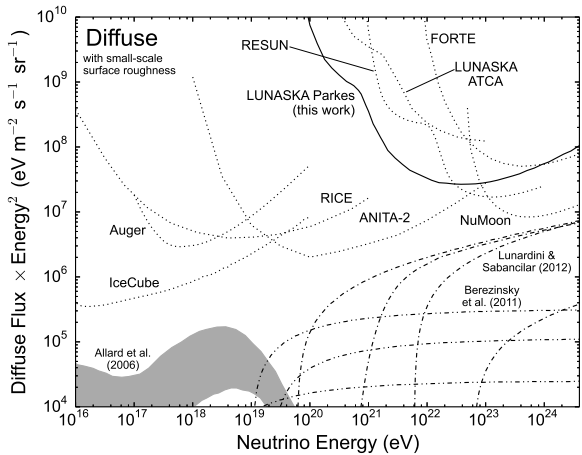}
 \caption{Model-independent limits to the diffuse neutrino flux from our experiment (solid) and others (dotted) compared to models for the expected neutrino flux, as in \figref{fig:iso_flux}, but with the limits for lunar radio experiments incorporating the effects of small-scale lunar surface roughness as described in \secref{sec:roughness}.  The limit for LUNASKA ATCA is as reported\gcitep{james2010}, and the limit for RESUN is derived from the reported limit\gcitep{jaeger2010}, which did not include small-scale surface roughness effects, by applying \eqnref{eqn:roughapspec}, adopting \mbox{$N_{\rm S} = 13$} and \mbox{$r = 0.70$} as done by \citet{james2010} for LUNASKA ATCA due to the similar detection scheme and radio frequency of these two experiments.  For NuMoon, \mbox{$N_{\rm S} \sim 1$}, and so we make no correction for small-scale surface roughness, displaying the originally-reported limit\gcitep{buitink2010}.}
 \label{fig:iso_flux_ssr}
\end{figure}

Based on the geometric aperture $A(E)$, we calculate the 90\%-confidence model-independent limit on a diffuse isotropic neutrino flux as defined by \citet{lehtinen2004}:
 \begin{equation}
  \frac{dF_{\rm iso}}{dE} < \frac{2.3}{E} \, \frac{1}{t_{\rm obs} \, A(E)} \label{eqn:lim_diffuse}
 \end{equation}
where $t_{\rm obs}$ is the effective observing time, and the factor 2.3 is based on the required confidence and the expected Poisson distribution of the number of detected neutrinos.  The resulting limit is shown in \figref{fig:iso_flux} alongside similar limits from other neutrino detection experiments; and again in \figref{fig:iso_flux_ssr}, with the effects of small-scale lunar surface roughness included.  Note that the lunar radio experiments (and FORTE), which are sensitive only at higher energies, also observed for less time, with a maximum of 200 hours for RESUN compared to 28.5 days for ANITA-2 and years for other non-lunar experiments.

The spectrum of the diffuse neutrino flux depends strongly on the UHECR origin model.  For bottom-up models, the UHE neutrino flux is expected to be dominated by neutrinos from GZK interactions.  The predicted flux depends strongly on the assumed source evolution model and on the composition of UHECRs.  We show in \figrefii{fig:iso_flux}{fig:iso_flux_ssr} a range of possible fluxes for different values of these parameters from \citet{allard2006}.  Recent results from the Pierre Auger Observatory indicate that there is a trend in the composition of UHECRs from protons to iron nuclei at energies above $10^{19}$ eV\gcitep{abraham2010b}, implying that the neutrino flux lies towards the bottom of this range.

For top-down models, there is expected to be an additional contribution to the UHE neutrino flux directly from the sources of UHECRs.  We show in \figrefii{fig:iso_flux}{fig:iso_flux_ssr} a selection of the more optimistic models not yet excluded by experiment\gcitep{berezinsky2011,lunardini2012}, which predict the neutrino fluxes resulting from the decay of weakly-interacting scalar particles called moduli, emitted from kinks and cusps in cosmic strings.  Due to the high fluxes predicted at higher neutrino energies, this family of models is more suited to being tested by lunar radio experiments.

Since an individual kink in a cosmic string produces a burst of UHE neutrinos, there is a possibility that two or more of these neutrinos may be detected near-simultaneously, which would be a clear signature of this origin mechanism\gcitep{lunardini2012}.  However, neutrinos which are too closely coincident would fail to be detected by our experiment.  For two near-simultaneous neutrinos interacting at different points on the lunar surface, typically separated by \mbox{$\sim 1$,000}~km, the times of arrival of the resulting Askaryan pulses will typically be separated by \mbox{$\gtrsim 3$}~ms.  With this separation, the first pulse would be detected and identified as originating from a lunar particle cascade, but the second pulse would arrive while the buffered data for the first event were being recorded, during which time our backend system is unable to respond to further triggers.  If the times of arrival of the neutrinos (and hence of the Askaryan pulses) were separated by \mbox{$\gtrsim 50$}~ms, both pulses would be recorded, but would be excluded as repeated impulsive RFI.  A pair of neutrinos could only both be detected if their separation exceeded the 10~s exclusion window of this anti-RFI cut.

\section{Directional neutrino sensitivity}
\label{sec:directional}

\begin{figure*}
 \centering
 \includegraphics[width=0.8\linewidth]{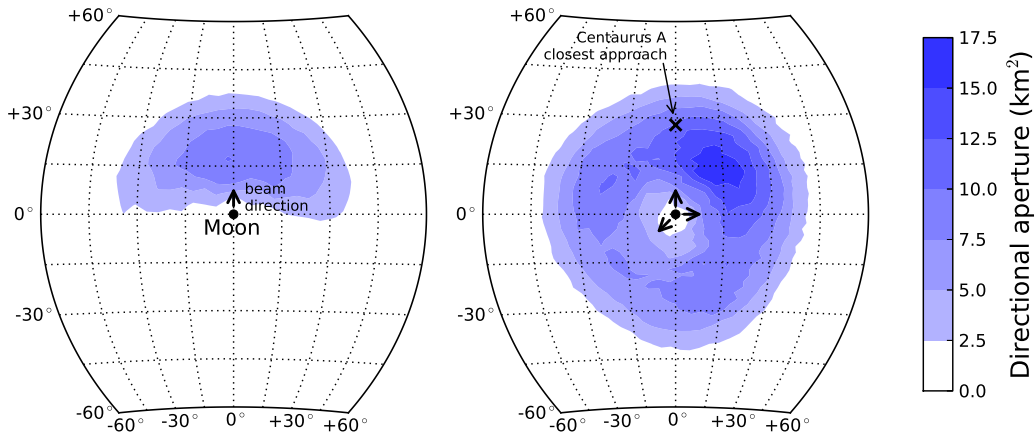}
 \caption{Directional aperture around the Moon for $10^{21}$~eV neutrinos, for a single limb beam (left) and for three beams as shown in \figref{fig:config} (right); beams are adjacent to the Moon in the directions indicated by the arrows.  The effects of anticoincidence rejection are included, but the effects of scattering due to small-scale lunar surface roughness are not; the roughness on the scale of the radio wavelength has a root-mean-square value of \mbox{$\sim 9\degree$}\gcitep{shepard1995}, and so the scattering may smooth the directional aperture up to roughly this scale.  The closest approach of \cena\ is 28\degree\ from the Moon in the direction shown.  Note that the maximum directional aperture is in the overlap between the regions from which a UHE neutrino could be detected by either of the limb beams.  The exposure to \cena\ could have been slightly increased if the limb beams were oriented to place it in the overlap region.}
 \label{fig:moon_ap_map}
\end{figure*}

\begin{figure*}
 \centering
 \includegraphics[width=0.9\linewidth]{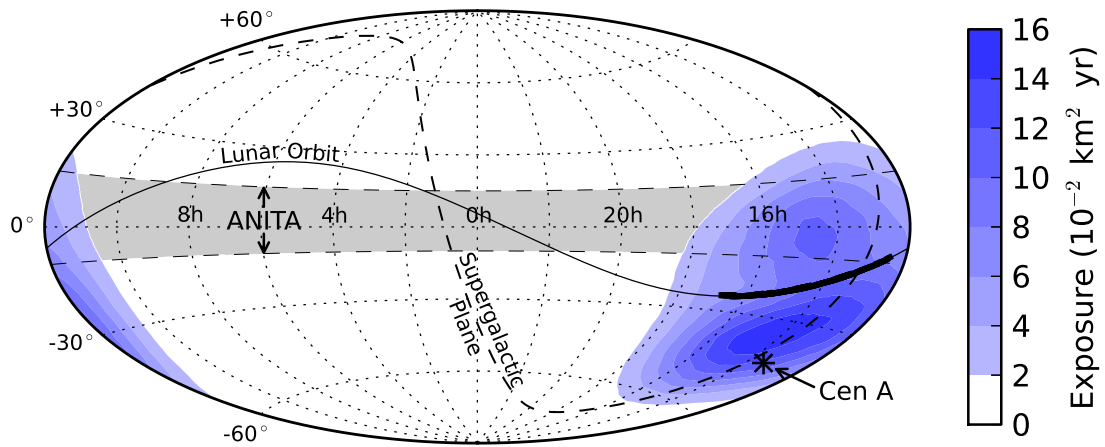}
 \caption{Directional exposure of our experiment, in equatorial coordinates, for $10^{21}$~eV neutrinos.  The effects of anticoincidence rejection are included, but the effects of scattering due to small-scale lunar surface roughness are not.  Observations were conducted while the Moon was on the section of its orbit marked with the thickened line, to maximize the exposure to \cena.  The ANITA experiment has substantially greater sensitivity than our experiment at this energy, but only in the declination range from $-10\degree$ to $+15\degree$, as shown.}
 \label{fig:iso_exp_map}
\end{figure*}

The simulation we use here to determine the neutrino aperture also calculates its directional dependence\gcitep{james2009f}.  The directional aperture around the Moon is shown in \figref{fig:moon_ap_map}.  We calculate the model-independent limit to a directional neutrino flux similarly to the diffuse case (\eqnref{eqn:lim_diffuse}), as 
 \begin{equation}
  \frac{dF_{\rm dir}}{dE} < \frac{2.3}{E} \, \frac{1}{\int \! dt \, A_{\rm dir}(E, \hat\Omega, t)}
 \end{equation}
where we integrate the directional aperture \mbox{$A_{\rm dir}(E, \hat\Omega, t)$} in the direction $\hat\Omega$ over the observing time.  This integral, the directional exposure, is shown in \figref{fig:iso_exp_map}.

% Contrary to the strategy described in \secref{sec:pointing}, it appears that the maximum directional aperture is in the overlap between the regions from which a UHE neutrino could be detected by either of the two limb beams, so the exposure to \cena\ would have been maximized by orienting the beams so that it was in this position.  However, \cena\ is typically at a sufficiently large angular distance from the Moon that this makes little difference.

\begin{figure}
 \centering
 \includegraphics[width=\linewidth]{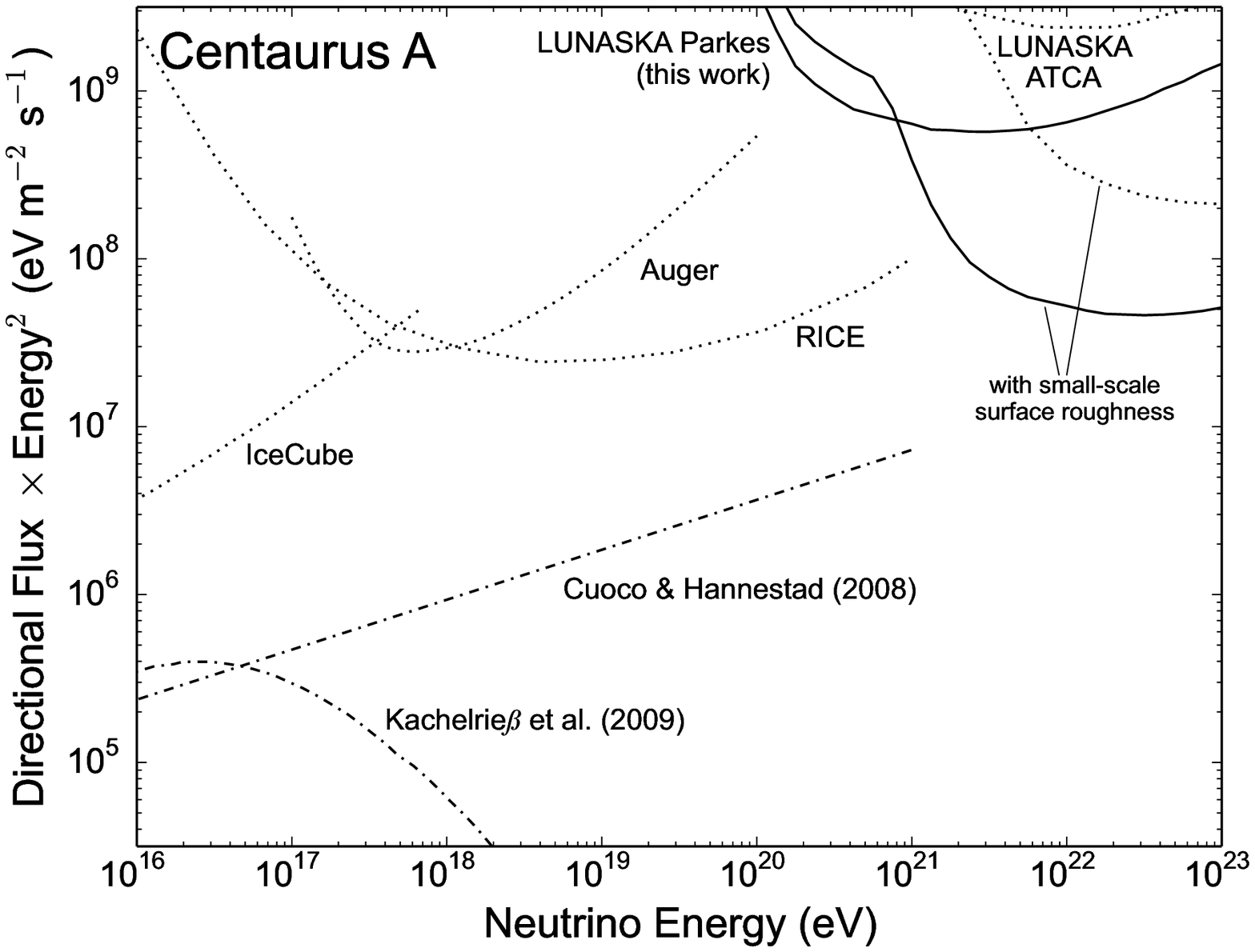}
 \caption{Model-independent limits to the neutrino flux from \cena\ from our experiment (solid), including the effects of anticoincidence rejection, and (dotted) for other experiments from \figref{fig:iso_flux} that have significant sensitivity in this direction, including the LUNASKA ATCA lunar radio experiment\gcitep{james2011}.  The limits for lunar radio experiments are shown both with (labeled) and without the effects of small-scale lunar surface roughness (see \secref{sec:roughness}).  The limits for IceCube\gcitep{aartsen2014} and the Pierre Auger Observatory\gcitep{abreu2011b,abreu2012} are derived from the reported model-dependent limits for \cena, with the energy dependence obtained for the former from the aperture for the appropriate declination range, and for the latter from the exposure to a diffuse flux.  For RICE\gcitep{kravchenko2012} the model-independent diffuse flux limit has been scaled based on the directional dependence of the sensitivity provided by \citet{besson2009}, as in \citet{james2011} but updated for the increased exposure.  Models for the neutrino flux from \cena\ are shown (dash-dotted) based on \citet{cuoco2008} and \citet{kachelriess2009}; see text for details.}
 \label{fig:cena_flux}
\end{figure}

\cena, a nearby AGN and closest FR-I type\gcitep{fanaroff1974} radio galaxy, is a potential source of UHECRs and hence also of UHE neutrinos.  As its distance \citep[3.8~Mpc;][]{harris2010} is less than the mean free path for GZK interactions of UHECR protons\gcitep{allard2009}, and the neutrino flux from these interactions is much reduced if the UHECRs are heavier nuclei, the neutrino flux from \cena\ (if it is a source of UHECRs) should be dominated by neutrinos produced by interactions in the source.  Several models for this flux exist, determining the relation between the UHECR and UHE neutrino fluxes based on the assumed conditions in the source region.  The limit that our experiment places on the neutrino flux from \cena\ is shown in \figref{fig:cena_flux} along with two such models.

\citet{cuoco2008} use a semi-analytic model to determine the expected neutrino flux from interactions of UHECRs, which are assumed to have an initial spectrum following a broken power-law,
% (with the exponent changing from -0.7 to -2.7 at $10^{17}$~eV),
 with ambient radiation and matter in the core region of \cena.  \citet{kachelriess2009} simulate the same interactions for a range of initial UHECR spectra, for conditions both in the core and in the jets of \cena, although the latter scenarios predict $\gamma$-ray spectra that are excluded by observations\gcitep{kachelriess2009b}; we show their scenario for UHECRs in the core region with a broken power-law spectrum,
% (with the exponent changing from -2 to -2.7 at $10^{18}$~eV),
 which gives the most optimistic neutrino flux.  Both models assume the UHECRs to be composed entirely of protons, and normalize the UHE neutrino flux to a UHECR flux based on the 2~UHECRs detected by the Pierre Auger Observatory originating from within 3\degree\ of \cena, with an exposure of 9,000 km$^2$~sr~yr\gcitep{abraham2008}.  In \figref{fig:cena_flux} we have rescaled the neutrino fluxes to a UHECR flux based on the larger-scale correlation seen in more recent data, with 13~UHECRs originating from within 18\degree\ of \cena, with an exposure of 20,370 km$^2$~sr~yr\gcitep{abreu2010}.  Note that, if these UHECRs are heavier nuclei rather than protons, which the greater magnetic deflection implied by this larger-scale correlation supports, the neutrino flux will be reduced.

The divergence at high energies between these two models, seen in \figref{fig:cena_flux}, is caused by the dependence of the UHE neutrino flux on the maximum energy of the initial UHECR spectrum.  \citeauthor{cuoco2008}\ assume that there is no such cut-off, as they were concerned with the neutrino flux at lower energies, where this assumption is unimportant; while \citeauthor{kachelriess2009}\ assume a maximum UHECR energy of $10^{20}$~eV.  The true value of the maximum UHECR energy depends on the magnetic field strength in the source region of \cena\gcitep{hillas1984}, which is not well constrained.

% Not including model of \citet{anchordoqui2011} as shown by \citet{guardincerri2011}.

\citet{koers2008} point out that if some fraction of the observed UHECRs are attributed to \cena, as assumed in the above models, there are implications for the ratio between the directional neutrino flux from \cena\ and the total diffuse neutrino flux, assuming that \cena\ is a typical source of both UHECRs and associated neutrinos.  The observed UHECR flux must originate from sources which, like \cena, are within the GZK horizon, whereas the associated neutrino flux may originate from sources at any distance.  For reasonable source evolution models, this implies that the majority of sources lie outside the GZK horizon and thus contribute only to the neutrino flux.  \cena, therefore, should be responsible for a larger fraction of the UHECR flux than of the associated neutrino flux.  This argument does not apply if \cena\ is an exceptionally efficient source of UHE neutrinos: for example, if UHECRs are produced by AGN generally, but giant radio galaxies such as \cena\ have a higher relative neutrino luminosity.

Based on the 2 out of 27 UHECRs associated with \cena\ in observational data from the Pierre Auger Observatory\gcitep{abraham2008}, the expected UHE neutrino flux from the entire sky is 200--5,000 times that for \cena\ alone, depending on the source evolution model, with the upper end of this range corresponding to the observed evolution of AGN\gcitep{koers2008}.  Rescaling to the 13 out of 69 UHECRs associated with \cena\ on a larger scale in a more recent dataset\gcitep{abreu2010}, this range becomes 80--2,000 times.  The exposure of this experiment to \cena\ relative to the mean whole-sky exposure varies with energy, but is typically \mbox{$\sim 10$} times greater.  Hence, targeting \cena\ increased the expected number of detections for this experiment by $1/8$th in the most optimistic case, assuming that \cena\ is not an exceptionally efficient source of UHE neutrinos.

We also consider the directional neutrino sensitivity for gamma-ray bursts (GRBs), which are transient sources of $\gamma$-rays that have been proposed as sources of UHECRs and neutrinos\gcitep{waxman1997}; the prospects for this have been limited by recent experimental results\gcitep{abbasi2012}, though not excluded\gcitep{hummer2012}.  Searching for neutrinos from GRBs is a tempting prospect because a coincidence with a GRB would increase the significance of a lunar-origin radio pulse in our data, reducing the effective noise threshold.  However, of the 9~GRBs in the catalogue maintained by the GRB Coordinates Network\gcitep{barthelmy2000} which occurred during our observations, none were close enough to the Moon for our experiment to be appreciably sensitive to them.

\section{Conclusion and summary}
\label{sec:conclusion}

We have established a limit on the UHE neutrino flux from our lunar radio experiment with the Parkes radio telescope.  Our experiment used a large collecting area and broad bandwidth, combined with a signal-processing strategy that came very close to the theoretical optimum performance (with 99.6\% efficiency), giving it a lower radio detection threshold than previous lunar radio experiments.  This translates to a limit that extends to lower energies than previous experiments of this type, establishing a tighter constraint on the diffuse neutrino flux for energies below $10^{21}$~eV.  Further improvement in this direction will require larger telescopes with broader bandwidth, such as the Square Kilometre Array: there is no substantial further improvement possible in the signal-processing performance.

Our limit at energies above $10^{22}$~eV, relevant for the detection of UHE neutrinos from some top-down models, does not represent an improvement over previous experiments.  The most obvious reason for this is the incomplete lunar coverage of the telescope beams in our experiment, reducing the neutrino aperture, which should also be remedied in future work.  However, the limits for previous experiments have been calculated with a range of sensitivity models, most of which neglect several effects which adversely affect the aperture.  In particular, the possibility of a lunar-origin Askaryan pulse being detected in multiple beams, and hence excluded as interference by an anticoincidence filter, significantly reduces the aperture at high neutrino energies in our experiment, and probably also in others.  Future analysis work for both past and future experiments should incorporate these effects.

There are two substantial sources of uncertainty in the reported neutrino flux limit.  The uncertainty in the neutrino-nucleon cross-section in the relevant energy range, under standard-model assumptions, is a factor of \mbox{$\sim 2$}, which translates almost linearly to a similar uncertainty in the flux limit.
 This cross-section is a matter of some theoretical interest, independent of the astrophysical origin of UHE neutrinos, but is degenerate with the neutrino flux in this experiment.
Our model for small-scale lunar surface roughness, which provides a rough upper bound on its effects, indicates that it improves the limit by a factor \mbox{$\sim 10$} for an $E^{-2}$ neutrino spectrum.  Taking the uncertainty to be the range between the limits with and without these effects, this corresponds to an uncertainty factor of \mbox{$\sim 3$} around their geometric mean.  This source of uncertainty, at least, is susceptible to further theoretical work, which should be a priority for this technique.

Our observations were planned to target the nearby AGN \cena, and achieved a sensitivity to neutrinos from this direction which is \mbox{$\sim 10\times$} greater than the mean sensitivity over the sky.  We have thus established a directional limit to the UHE neutrino flux from \cena\ which is the lowest published limit for neutrino energies above $10^{21}$~eV.  However, this targeting makes only a modest contribution to the expected event rate, unless \cena\ is an exceptionally efficient neutrino source.  The targeting strategy may be better justified in terms of directional sensitivity to UHECRs, of which \cena\ potentially contributes a larger fraction of the all-sky flux.

\begin{acknowledgments}
 The Parkes radio telescope is part of the Australia Telescope which is funded by the Commonwealth of Australia for operation as a National Facility managed by CSIRO.  Some data used in this study were acquired as part of NASA's Earth Science Data Systems and archived and distributed by the Crustal Dynamics Data Information System (CDDIS).  This research was supported by the Australian Research Council's Discovery Project funding scheme (project number DP0881006).  JDB acknowledges support from ERC-StG 307215 (LODESTONE).
\end{acknowledgments}

\bibliographystyle{apsrev}
\bibliography{all}

\begin{thebibliography}{58}
\expandafter\ifx\csname natexlab\endcsname\relax\def\natexlab#1{#1}\fi
\expandafter\ifx\csname bibnamefont\endcsname\relax
  \def\bibnamefont#1{#1}\fi
\expandafter\ifx\csname bibfnamefont\endcsname\relax
  \def\bibfnamefont#1{#1}\fi
\expandafter\ifx\csname citenamefont\endcsname\relax
  \def\citenamefont#1{#1}\fi
\expandafter\ifx\csname url\endcsname\relax
  \def\url#1{\texttt{#1}}\fi
\expandafter\ifx\csname urlprefix\endcsname\relax\def\urlprefix{URL }\fi
\providecommand{\bibinfo}[2]{#2}
\providecommand{\eprint}[2][]{\url{#2}}

\bibitem[{\citenamefont{{Greisen}}(1966)}]{greisen1966}
\bibinfo{author}{\bibfnamefont{K.}~\bibnamefont{{Greisen}}},
  \bibinfo{journal}{\prl{}} \textbf{\bibinfo{volume}{16}}, \bibinfo{pages}{748}
  (\bibinfo{year}{1966}).

\bibitem[{\citenamefont{{Zatsepin} and {Kuzmin}}(1966)}]{zatsepin1966}
\bibinfo{author}{\bibfnamefont{G.~T.} \bibnamefont{{Zatsepin}}}
  \bibnamefont{and} \bibinfo{author}{\bibfnamefont{V.~A.}
  \bibnamefont{{Kuzmin}}}, \bibinfo{journal}{\spjetpl{}}
  \textbf{\bibinfo{volume}{4}}, \bibinfo{pages}{78} (\bibinfo{year}{1966}).

\bibitem[{\citenamefont{{Abraham}
  et~al.}(2010{\natexlab{a}})\citenamefont{{Abraham}, {Abreu}, {Aglietta},
  {Ahn}, {Allard}, {Allen}, {Alvarez-Mu{\~n}iz}, {Ambrosio}, {Anchordoqui},
  {Andringa} et~al.}}]{abraham2010}
\bibinfo{author}{\bibfnamefont{J.}~\bibnamefont{{Abraham}}},
  \bibinfo{author}{\bibfnamefont{P.}~\bibnamefont{{Abreu}}},
  \bibinfo{author}{\bibfnamefont{M.}~\bibnamefont{{Aglietta}}},
  \bibinfo{author}{\bibfnamefont{E.~J.} \bibnamefont{{Ahn}}},
  \bibinfo{author}{\bibfnamefont{D.}~\bibnamefont{{Allard}}},
  \bibinfo{author}{\bibfnamefont{J.}~\bibnamefont{{Allen}}},
  \bibinfo{author}{\bibfnamefont{J.}~\bibnamefont{{Alvarez-Mu{\~n}iz}}},
  \bibinfo{author}{\bibfnamefont{M.}~\bibnamefont{{Ambrosio}}},
  \bibinfo{author}{\bibfnamefont{L.}~\bibnamefont{{Anchordoqui}}},
  \bibinfo{author}{\bibfnamefont{S.}~\bibnamefont{{Andringa}}},
  \bibnamefont{et~al.} (\bibinfo{collaboration}{Pierre Auger Collaboration}),
  \bibinfo{journal}{\plb{}} \textbf{\bibinfo{volume}{685}},
  \bibinfo{pages}{239} (\bibinfo{year}{2010}{\natexlab{a}}).

\bibitem[{\citenamefont{{Ackermann} et~al.}(2013)\citenamefont{{Ackermann},
  {Ajello}, {Allafort}, {Baldini}, {Ballet}, {Barbiellini}, {Baring},
  {Bastieri}, {Bechtol}, {Bellazzini} et~al.}}]{ackermann2013}
\bibinfo{author}{\bibfnamefont{M.}~\bibnamefont{{Ackermann}}},
  \bibinfo{author}{\bibfnamefont{M.}~\bibnamefont{{Ajello}}},
  \bibinfo{author}{\bibfnamefont{A.}~\bibnamefont{{Allafort}}},
  \bibinfo{author}{\bibfnamefont{L.}~\bibnamefont{{Baldini}}},
  \bibinfo{author}{\bibfnamefont{J.}~\bibnamefont{{Ballet}}},
  \bibinfo{author}{\bibfnamefont{G.}~\bibnamefont{{Barbiellini}}},
  \bibinfo{author}{\bibfnamefont{M.~G.} \bibnamefont{{Baring}}},
  \bibinfo{author}{\bibfnamefont{D.}~\bibnamefont{{Bastieri}}},
  \bibinfo{author}{\bibfnamefont{K.}~\bibnamefont{{Bechtol}}},
  \bibinfo{author}{\bibfnamefont{R.}~\bibnamefont{{Bellazzini}}},
  \bibnamefont{et~al.}, \bibinfo{journal}{\sci{}}
  \textbf{\bibinfo{volume}{339}}, \bibinfo{pages}{807} (\bibinfo{year}{2013}).

\bibitem[{\citenamefont{{Abraham} et~al.}(2008)\citenamefont{{Abraham},
  {Abreu}, {Aglietta}, {Aguirre}, {Allard}, {Allekotte}, {Allen}, {Allison},
  {Alvarez-Mu{\~n}iz}, {Ambrosio} et~al.}}]{abraham2008}
\bibinfo{author}{\bibfnamefont{J.}~\bibnamefont{{Abraham}}},
  \bibinfo{author}{\bibfnamefont{P.}~\bibnamefont{{Abreu}}},
  \bibinfo{author}{\bibfnamefont{M.}~\bibnamefont{{Aglietta}}},
  \bibinfo{author}{\bibfnamefont{C.}~\bibnamefont{{Aguirre}}},
  \bibinfo{author}{\bibfnamefont{D.}~\bibnamefont{{Allard}}},
  \bibinfo{author}{\bibfnamefont{I.}~\bibnamefont{{Allekotte}}},
  \bibinfo{author}{\bibfnamefont{J.}~\bibnamefont{{Allen}}},
  \bibinfo{author}{\bibfnamefont{P.}~\bibnamefont{{Allison}}},
  \bibinfo{author}{\bibfnamefont{J.}~\bibnamefont{{Alvarez-Mu{\~n}iz}}},
  \bibinfo{author}{\bibfnamefont{M.}~\bibnamefont{{Ambrosio}}},
  \bibnamefont{et~al.} (\bibinfo{collaboration}{Pierre Auger Collaboration}),
  \bibinfo{journal}{\app{}} \textbf{\bibinfo{volume}{29}}, \bibinfo{pages}{188}
  (\bibinfo{year}{2008}).

\bibitem[{\citenamefont{{Abreu} et~al.}(2010)\citenamefont{{Abreu}, {Aglietta},
  {Ahn}, {Allard}, {Allekotte}, {Allen}, {Alvarez Castillo},
  {Alvarez-Mu{\~n}iz}, {Ambrosio}, {Aminaei} et~al.}}]{abreu2010}
\bibinfo{author}{\bibfnamefont{P.}~\bibnamefont{{Abreu}}},
  \bibinfo{author}{\bibfnamefont{M.}~\bibnamefont{{Aglietta}}},
  \bibinfo{author}{\bibfnamefont{E.~J.} \bibnamefont{{Ahn}}},
  \bibinfo{author}{\bibfnamefont{D.}~\bibnamefont{{Allard}}},
  \bibinfo{author}{\bibfnamefont{I.}~\bibnamefont{{Allekotte}}},
  \bibinfo{author}{\bibfnamefont{J.}~\bibnamefont{{Allen}}},
  \bibinfo{author}{\bibfnamefont{J.}~\bibnamefont{{Alvarez Castillo}}},
  \bibinfo{author}{\bibfnamefont{J.}~\bibnamefont{{Alvarez-Mu{\~n}iz}}},
  \bibinfo{author}{\bibfnamefont{M.}~\bibnamefont{{Ambrosio}}},
  \bibinfo{author}{\bibfnamefont{A.}~\bibnamefont{{Aminaei}}},
  \bibnamefont{et~al.} (\bibinfo{collaboration}{Pierre Auger Collaboration}),
  \bibinfo{journal}{\app{}} \textbf{\bibinfo{volume}{34}}, \bibinfo{pages}{314}
  (\bibinfo{year}{2010}).

\bibitem[{\citenamefont{{Protheroe}}(2004)}]{protheroe2004}
\bibinfo{author}{\bibfnamefont{R.~J.} \bibnamefont{{Protheroe}}},
  \bibinfo{journal}{\app{}} \textbf{\bibinfo{volume}{21}}, \bibinfo{pages}{415}
  (\bibinfo{year}{2004}).

\bibitem[{\citenamefont{{Askaryan}}(1962)}]{askaryan1962}
\bibinfo{author}{\bibfnamefont{G.~A.} \bibnamefont{{Askaryan}}},
  \bibinfo{journal}{\spjetp{}} \textbf{\bibinfo{volume}{14}},
  \bibinfo{pages}{441} (\bibinfo{year}{1962}).

\bibitem[{\citenamefont{{Saltzberg} et~al.}(2001)\citenamefont{{Saltzberg},
  {Gorham}, {Walz}, {Field}, {Iverson}, {Odian}, {Resch}, {Schoessow}, and
  {Williams}}}]{saltzberg2001}
\bibinfo{author}{\bibfnamefont{D.}~\bibnamefont{{Saltzberg}}},
  \bibinfo{author}{\bibfnamefont{P.}~\bibnamefont{{Gorham}}},
  \bibinfo{author}{\bibfnamefont{D.}~\bibnamefont{{Walz}}},
  \bibinfo{author}{\bibfnamefont{C.}~\bibnamefont{{Field}}},
  \bibinfo{author}{\bibfnamefont{R.}~\bibnamefont{{Iverson}}},
  \bibinfo{author}{\bibfnamefont{A.}~\bibnamefont{{Odian}}},
  \bibinfo{author}{\bibfnamefont{G.}~\bibnamefont{{Resch}}},
  \bibinfo{author}{\bibfnamefont{P.}~\bibnamefont{{Schoessow}}},
  \bibnamefont{and}
  \bibinfo{author}{\bibfnamefont{D.}~\bibnamefont{{Williams}}},
  \bibinfo{journal}{\prl{}} \textbf{\bibinfo{volume}{86}},
  \bibinfo{pages}{2802} (\bibinfo{year}{2001}).

\bibitem[{\citenamefont{{Gorham} et~al.}(2005)\citenamefont{{Gorham},
  {Saltzberg}, {Field}, {Guillian}, {Milin{\v c}i{\'c}}, {Mio{\v c}inovi{\'c}},
  {Walz}, and {Williams}}}]{gorham2005}
\bibinfo{author}{\bibfnamefont{P.~W.} \bibnamefont{{Gorham}}},
  \bibinfo{author}{\bibfnamefont{D.}~\bibnamefont{{Saltzberg}}},
  \bibinfo{author}{\bibfnamefont{R.~C.} \bibnamefont{{Field}}},
  \bibinfo{author}{\bibfnamefont{E.}~\bibnamefont{{Guillian}}},
  \bibinfo{author}{\bibfnamefont{R.}~\bibnamefont{{Milin{\v c}i{\'c}}}},
  \bibinfo{author}{\bibfnamefont{P.}~\bibnamefont{{Mio{\v c}inovi{\'c}}}},
  \bibinfo{author}{\bibfnamefont{D.}~\bibnamefont{{Walz}}}, \bibnamefont{and}
  \bibinfo{author}{\bibfnamefont{D.}~\bibnamefont{{Williams}}},
  \bibinfo{journal}{\prd{}} \textbf{\bibinfo{volume}{72}},
  \bibinfo{eid}{023002} (\bibinfo{year}{2005}).

\bibitem[{\citenamefont{{Gorham} et~al.}(2007)\citenamefont{{Gorham},
  {Barwick}, {Beatty}, {Besson}, {Binns}, {Chen}, {Chen}, {Clem}, {Connolly},
  {Dowkontt} et~al.}}]{gorham2007}
\bibinfo{author}{\bibfnamefont{P.~W.} \bibnamefont{{Gorham}}},
  \bibinfo{author}{\bibfnamefont{S.~W.} \bibnamefont{{Barwick}}},
  \bibinfo{author}{\bibfnamefont{J.~J.} \bibnamefont{{Beatty}}},
  \bibinfo{author}{\bibfnamefont{D.~Z.} \bibnamefont{{Besson}}},
  \bibinfo{author}{\bibfnamefont{W.~R.} \bibnamefont{{Binns}}},
  \bibinfo{author}{\bibfnamefont{C.}~\bibnamefont{{Chen}}},
  \bibinfo{author}{\bibfnamefont{P.}~\bibnamefont{{Chen}}},
  \bibinfo{author}{\bibfnamefont{J.~M.} \bibnamefont{{Clem}}},
  \bibinfo{author}{\bibfnamefont{A.}~\bibnamefont{{Connolly}}},
  \bibinfo{author}{\bibfnamefont{P.~F.} \bibnamefont{{Dowkontt}}},
  \bibnamefont{et~al.} (\bibinfo{collaboration}{{ANITA Collaboration}}),
  \bibinfo{journal}{\prl{}} \textbf{\bibinfo{volume}{99}},
  \bibinfo{eid}{171101} (\bibinfo{year}{2007}).

\bibitem[{\citenamefont{{Kravchenko} et~al.}(2006)\citenamefont{{Kravchenko},
  {Cooley}, {Hussain}, {Seckel}, {Wahrlich}, {Adams}, {Churchwell}, {Harris},
  {Seunarine}, {Bean} et~al.}}]{kravchenko2006}
\bibinfo{author}{\bibfnamefont{I.}~\bibnamefont{{Kravchenko}}},
  \bibinfo{author}{\bibfnamefont{C.}~\bibnamefont{{Cooley}}},
  \bibinfo{author}{\bibfnamefont{S.}~\bibnamefont{{Hussain}}},
  \bibinfo{author}{\bibfnamefont{D.}~\bibnamefont{{Seckel}}},
  \bibinfo{author}{\bibfnamefont{P.}~\bibnamefont{{Wahrlich}}},
  \bibinfo{author}{\bibfnamefont{J.}~\bibnamefont{{Adams}}},
  \bibinfo{author}{\bibfnamefont{S.}~\bibnamefont{{Churchwell}}},
  \bibinfo{author}{\bibfnamefont{P.}~\bibnamefont{{Harris}}},
  \bibinfo{author}{\bibfnamefont{S.}~\bibnamefont{{Seunarine}}},
  \bibinfo{author}{\bibfnamefont{A.}~\bibnamefont{{Bean}}},
  \bibnamefont{et~al.}, \bibinfo{journal}{\prd{}}
  \textbf{\bibinfo{volume}{73}}, \bibinfo{pages}{082002}
  (\bibinfo{year}{2006}).

\bibitem[{\citenamefont{{Gorham} et~al.}(2009)\citenamefont{{Gorham},
  {Allison}, {Barwick}, {Beatty}, {Besson}, {Binns}, {Chen}, {Chen}, {Clem},
  {Connolly} et~al.}}]{gorham2009b}
\bibinfo{author}{\bibfnamefont{P.~W.} \bibnamefont{{Gorham}}},
  \bibinfo{author}{\bibfnamefont{P.}~\bibnamefont{{Allison}}},
  \bibinfo{author}{\bibfnamefont{S.~W.} \bibnamefont{{Barwick}}},
  \bibinfo{author}{\bibfnamefont{J.~J.} \bibnamefont{{Beatty}}},
  \bibinfo{author}{\bibfnamefont{D.~Z.} \bibnamefont{{Besson}}},
  \bibinfo{author}{\bibfnamefont{W.~R.} \bibnamefont{{Binns}}},
  \bibinfo{author}{\bibfnamefont{C.}~\bibnamefont{{Chen}}},
  \bibinfo{author}{\bibfnamefont{P.}~\bibnamefont{{Chen}}},
  \bibinfo{author}{\bibfnamefont{J.~M.} \bibnamefont{{Clem}}},
  \bibinfo{author}{\bibfnamefont{A.}~\bibnamefont{{Connolly}}},
  \bibnamefont{et~al.} (\bibinfo{collaboration}{{ANITA Collaboration}}),
  \bibinfo{journal}{\app{}} \textbf{\bibinfo{volume}{32}}, \bibinfo{pages}{10}
  (\bibinfo{year}{2009}).

\bibitem[{\citenamefont{{Lehtinen} et~al.}(2004)\citenamefont{{Lehtinen},
  {Gorham}, {Jacobson}, and {Roussel-Dupr{\'e}}}}]{lehtinen2004}
\bibinfo{author}{\bibfnamefont{N.~G.} \bibnamefont{{Lehtinen}}},
  \bibinfo{author}{\bibfnamefont{P.~W.} \bibnamefont{{Gorham}}},
  \bibinfo{author}{\bibfnamefont{A.~R.} \bibnamefont{{Jacobson}}},
  \bibnamefont{and} \bibinfo{author}{\bibfnamefont{R.~A.}
  \bibnamefont{{Roussel-Dupr{\'e}}}}, \bibinfo{journal}{\prd{}}
  \textbf{\bibinfo{volume}{69}}, \bibinfo{pages}{013008}
  (\bibinfo{year}{2004}).

\bibitem[{\citenamefont{{Dagkesamanskii} and
  {Zheleznykh}}(1989)}]{dagkesamanskii1989}
\bibinfo{author}{\bibfnamefont{R.~D.} \bibnamefont{{Dagkesamanskii}}}
  \bibnamefont{and} \bibinfo{author}{\bibfnamefont{I.~M.}
  \bibnamefont{{Zheleznykh}}}, \bibinfo{journal}{\spjetpl{}}
  \textbf{\bibinfo{volume}{50}}, \bibinfo{pages}{259} (\bibinfo{year}{1989}).

\bibitem[{\citenamefont{{Protheroe} and {Stanev}}(1996)}]{protheroe1996b}
\bibinfo{author}{\bibfnamefont{R.~J.} \bibnamefont{{Protheroe}}}
  \bibnamefont{and} \bibinfo{author}{\bibfnamefont{T.}~\bibnamefont{{Stanev}}},
  \bibinfo{journal}{\prl{}} \textbf{\bibinfo{volume}{77}},
  \bibinfo{pages}{3708} (\bibinfo{year}{1996}).

\bibitem[{\citenamefont{{Berezinsky} et~al.}(2011)\citenamefont{{Berezinsky},
  {Sabancilar}, and {Vilenkin}}}]{berezinsky2011}
\bibinfo{author}{\bibfnamefont{V.}~\bibnamefont{{Berezinsky}}},
  \bibinfo{author}{\bibfnamefont{E.}~\bibnamefont{{Sabancilar}}},
  \bibnamefont{and}
  \bibinfo{author}{\bibfnamefont{A.}~\bibnamefont{{Vilenkin}}},
  \bibinfo{journal}{\prd{}} \textbf{\bibinfo{volume}{84}},
  \bibinfo{eid}{085006} (\bibinfo{year}{2011}).

\bibitem[{\citenamefont{{Lunardini} and {Sabancilar}}(2012)}]{lunardini2012}
\bibinfo{author}{\bibfnamefont{C.}~\bibnamefont{{Lunardini}}} \bibnamefont{and}
  \bibinfo{author}{\bibfnamefont{E.}~\bibnamefont{{Sabancilar}}},
  \bibinfo{journal}{\prd{}} \textbf{\bibinfo{volume}{86}},
  \bibinfo{eid}{085008} (\bibinfo{year}{2012}).

\bibitem[{\citenamefont{{ter Veen} et~al.}(2010)\citenamefont{{ter Veen},
  {Buitink}, {Falcke}, {James}, {Mevius}, {Scholten}, {Singh}, {Stappers}, and
  {de Vries}}}]{terveen2010}
\bibinfo{author}{\bibfnamefont{S.}~\bibnamefont{{ter Veen}}},
  \bibinfo{author}{\bibfnamefont{S.}~\bibnamefont{{Buitink}}},
  \bibinfo{author}{\bibfnamefont{H.}~\bibnamefont{{Falcke}}},
  \bibinfo{author}{\bibfnamefont{C.~W.} \bibnamefont{{James}}},
  \bibinfo{author}{\bibfnamefont{M.}~\bibnamefont{{Mevius}}},
  \bibinfo{author}{\bibfnamefont{O.}~\bibnamefont{{Scholten}}},
  \bibinfo{author}{\bibfnamefont{K.}~\bibnamefont{{Singh}}},
  \bibinfo{author}{\bibfnamefont{B.}~\bibnamefont{{Stappers}}},
  \bibnamefont{and} \bibinfo{author}{\bibfnamefont{K.~D.} \bibnamefont{{de
  Vries}}}, \bibinfo{journal}{\prd{}} \textbf{\bibinfo{volume}{82}},
  \bibinfo{pages}{103014} (\bibinfo{year}{2010}).

\bibitem[{\citenamefont{{Jeong} et~al.}(2012)\citenamefont{{Jeong}, {Reno}, and
  {Sarcevic}}}]{jeong2012}
\bibinfo{author}{\bibfnamefont{Y.~S.} \bibnamefont{{Jeong}}},
  \bibinfo{author}{\bibfnamefont{M.~H.} \bibnamefont{{Reno}}},
  \bibnamefont{and}
  \bibinfo{author}{\bibfnamefont{I.}~\bibnamefont{{Sarcevic}}},
  \bibinfo{journal}{\app{}} \textbf{\bibinfo{volume}{35}}, \bibinfo{pages}{383}
  (\bibinfo{year}{2012}).

\bibitem[{\citenamefont{{Carilli} and {Rawlings}}(2004)}]{carilli2004}
\bibinfo{author}{\bibfnamefont{C.~L.} \bibnamefont{{Carilli}}}
  \bibnamefont{and}
  \bibinfo{author}{\bibfnamefont{S.}~\bibnamefont{{Rawlings}}},
  \bibinfo{journal}{\nar{}} \textbf{\bibinfo{volume}{48}}, \bibinfo{pages}{979}
  (\bibinfo{year}{2004}).

\bibitem[{\citenamefont{{Bray} et~al.}(2015)\citenamefont{{Bray}, {Ekers},
  {Roberts}, {Reynolds}, {James}, {Phillips}, {Protheroe}, {McFadden}, and
  {Aartsen}}}]{bray2014a}
\bibinfo{author}{\bibfnamefont{J.~D.} \bibnamefont{{Bray}}},
  \bibinfo{author}{\bibfnamefont{R.~D.} \bibnamefont{{Ekers}}},
  \bibinfo{author}{\bibfnamefont{P.}~\bibnamefont{{Roberts}}},
  \bibinfo{author}{\bibfnamefont{J.~E.} \bibnamefont{{Reynolds}}},
  \bibinfo{author}{\bibfnamefont{C.~W.} \bibnamefont{{James}}},
  \bibinfo{author}{\bibfnamefont{C.~J.} \bibnamefont{{Phillips}}},
  \bibinfo{author}{\bibfnamefont{R.~J.} \bibnamefont{{Protheroe}}},
  \bibinfo{author}{\bibfnamefont{R.~A.} \bibnamefont{{McFadden}}},
  \bibnamefont{and} \bibinfo{author}{\bibfnamefont{M.~G.}
  \bibnamefont{{Aartsen}}}, \bibinfo{journal}{\app{}}
  \textbf{\bibinfo{volume}{65}}, \bibinfo{pages}{22} (\bibinfo{year}{2015}).

\bibitem[{\citenamefont{{Staveley-Smith}
  et~al.}(1996)\citenamefont{{Staveley-Smith}, {Wilson}, {Bird}, {Disney},
  {Ekers}, {Freeman}, {Haynes}, {Sinclair}, {Vaile}, {Webster}
  et~al.}}]{staveley-smith1996}
\bibinfo{author}{\bibfnamefont{L.}~\bibnamefont{{Staveley-Smith}}},
  \bibinfo{author}{\bibfnamefont{W.~E.} \bibnamefont{{Wilson}}},
  \bibinfo{author}{\bibfnamefont{T.~S.} \bibnamefont{{Bird}}},
  \bibinfo{author}{\bibfnamefont{M.~J.} \bibnamefont{{Disney}}},
  \bibinfo{author}{\bibfnamefont{R.~D.} \bibnamefont{{Ekers}}},
  \bibinfo{author}{\bibfnamefont{K.~C.} \bibnamefont{{Freeman}}},
  \bibinfo{author}{\bibfnamefont{R.~F.} \bibnamefont{{Haynes}}},
  \bibinfo{author}{\bibfnamefont{M.~W.} \bibnamefont{{Sinclair}}},
  \bibinfo{author}{\bibfnamefont{R.~A.} \bibnamefont{{Vaile}}},
  \bibinfo{author}{\bibfnamefont{R.~L.} \bibnamefont{{Webster}}},
  \bibnamefont{et~al.}, \bibinfo{journal}{\pasa{}}
  \textbf{\bibinfo{volume}{13}}, \bibinfo{pages}{243} (\bibinfo{year}{1996}).

\bibitem[{\citenamefont{{James} and
  {Protheroe}}(2009{\natexlab{a}})}]{james2009f}
\bibinfo{author}{\bibfnamefont{C.~W.} \bibnamefont{{James}}} \bibnamefont{and}
  \bibinfo{author}{\bibfnamefont{R.~J.} \bibnamefont{{Protheroe}}},
  \bibinfo{journal}{\app{}} \textbf{\bibinfo{volume}{31}}, \bibinfo{pages}{392}
  (\bibinfo{year}{2009}{\natexlab{a}}).

\bibitem[{\citenamefont{{James} et~al.}(2010)\citenamefont{{James}, {Ekers},
  {Alvarez-Mu{\~n}iz}, {Bray}, {McFadden}, {Phillips}, {Protheroe}, and
  {Roberts}}}]{james2010}
\bibinfo{author}{\bibfnamefont{C.~W.} \bibnamefont{{James}}},
  \bibinfo{author}{\bibfnamefont{R.~D.} \bibnamefont{{Ekers}}},
  \bibinfo{author}{\bibfnamefont{J.}~\bibnamefont{{Alvarez-Mu{\~n}iz}}},
  \bibinfo{author}{\bibfnamefont{J.~D.} \bibnamefont{{Bray}}},
  \bibinfo{author}{\bibfnamefont{R.~A.} \bibnamefont{{McFadden}}},
  \bibinfo{author}{\bibfnamefont{C.~J.} \bibnamefont{{Phillips}}},
  \bibinfo{author}{\bibfnamefont{R.~J.} \bibnamefont{{Protheroe}}},
  \bibnamefont{and}
  \bibinfo{author}{\bibfnamefont{P.}~\bibnamefont{{Roberts}}},
  \bibinfo{journal}{\prd{}} \textbf{\bibinfo{volume}{81}},
  \bibinfo{pages}{042003} (\bibinfo{year}{2010}).

\bibitem[{\citenamefont{{Bray} et~al.}(2013)\citenamefont{{Bray}, {Ekers}, and
  {Roberts}}}]{bray2012}
\bibinfo{author}{\bibfnamefont{J.~D.} \bibnamefont{{Bray}}},
  \bibinfo{author}{\bibfnamefont{R.~D.} \bibnamefont{{Ekers}}},
  \bibnamefont{and}
  \bibinfo{author}{\bibfnamefont{P.}~\bibnamefont{{Roberts}}},
  \bibinfo{journal}{\expa{}} \textbf{\bibinfo{volume}{36}},
  \bibinfo{pages}{155} (\bibinfo{year}{2013}), ISSN \bibinfo{issn}{0922-6435}.

\bibitem[{\citenamefont{{Moffat}}(1972)}]{moffat1972}
\bibinfo{author}{\bibfnamefont{P.~H.} \bibnamefont{{Moffat}}},
  \bibinfo{journal}{\mnras{}} \textbf{\bibinfo{volume}{160}},
  \bibinfo{pages}{139} (\bibinfo{year}{1972}).

\bibitem[{\citenamefont{{James} and
  {Protheroe}}(2009{\natexlab{b}})}]{james2009b}
\bibinfo{author}{\bibfnamefont{C.~W.} \bibnamefont{{James}}} \bibnamefont{and}
  \bibinfo{author}{\bibfnamefont{R.~J.} \bibnamefont{{Protheroe}}},
  \bibinfo{journal}{\app{}} \textbf{\bibinfo{volume}{30}}, \bibinfo{pages}{318}
  (\bibinfo{year}{2009}{\natexlab{b}}).

\bibitem[{\citenamefont{{Staveley-Smith}}(2009)}]{staveley-smith2009}
\bibinfo{author}{\bibfnamefont{L.}~\bibnamefont{{Staveley-Smith}}},
  \bibinfo{howpublished}{\privcom{}} (\bibinfo{year}{2009}).

\bibitem[{\citenamefont{{Gandhi} et~al.}(1998)\citenamefont{{Gandhi}, {Quigg},
  {Reno}, and {Sarcevic}}}]{gandhi1998}
\bibinfo{author}{\bibfnamefont{R.}~\bibnamefont{{Gandhi}}},
  \bibinfo{author}{\bibfnamefont{C.}~\bibnamefont{{Quigg}}},
  \bibinfo{author}{\bibfnamefont{M.~H.} \bibnamefont{{Reno}}},
  \bibnamefont{and}
  \bibinfo{author}{\bibfnamefont{I.}~\bibnamefont{{Sarcevic}}},
  \bibinfo{journal}{\prd{}} \textbf{\bibinfo{volume}{58}},
  \bibinfo{pages}{093009} (\bibinfo{year}{1998}).

\bibitem[{\citenamefont{{James} et~al.}(2011)\citenamefont{{James},
  {Protheroe}, {Ekers}, {Alvarez-Mu{\~n}iz}, {McFadden}, {Phillips}, {Roberts},
  and {Bray}}}]{james2011}
\bibinfo{author}{\bibfnamefont{C.~W.} \bibnamefont{{James}}},
  \bibinfo{author}{\bibfnamefont{R.~J.} \bibnamefont{{Protheroe}}},
  \bibinfo{author}{\bibfnamefont{R.~D.} \bibnamefont{{Ekers}}},
  \bibinfo{author}{\bibfnamefont{J.}~\bibnamefont{{Alvarez-Mu{\~n}iz}}},
  \bibinfo{author}{\bibfnamefont{R.~A.} \bibnamefont{{McFadden}}},
  \bibinfo{author}{\bibfnamefont{C.~J.} \bibnamefont{{Phillips}}},
  \bibinfo{author}{\bibfnamefont{P.}~\bibnamefont{{Roberts}}},
  \bibnamefont{and} \bibinfo{author}{\bibfnamefont{J.~D.}
  \bibnamefont{{Bray}}}, \bibinfo{journal}{\mnras{}}
  \textbf{\bibinfo{volume}{410}}, \bibinfo{pages}{885} (\bibinfo{year}{2011}).

\bibitem[{\citenamefont{{Buitink} et~al.}(2010)\citenamefont{{Buitink},
  {Scholten}, {Bacelar}, {Braun}, {de Bruyn}, {Falcke}, {Singh}, {Stappers},
  {Strom}, and {Yahyaoui}}}]{buitink2010}
\bibinfo{author}{\bibfnamefont{S.}~\bibnamefont{{Buitink}}},
  \bibinfo{author}{\bibfnamefont{O.}~\bibnamefont{{Scholten}}},
  \bibinfo{author}{\bibfnamefont{J.}~\bibnamefont{{Bacelar}}},
  \bibinfo{author}{\bibfnamefont{R.}~\bibnamefont{{Braun}}},
  \bibinfo{author}{\bibfnamefont{A.~G.} \bibnamefont{{de Bruyn}}},
  \bibinfo{author}{\bibfnamefont{H.}~\bibnamefont{{Falcke}}},
  \bibinfo{author}{\bibfnamefont{K.}~\bibnamefont{{Singh}}},
  \bibinfo{author}{\bibfnamefont{B.}~\bibnamefont{{Stappers}}},
  \bibinfo{author}{\bibfnamefont{R.~G.} \bibnamefont{{Strom}}},
  \bibnamefont{and} \bibinfo{author}{\bibfnamefont{R.~A.}
  \bibnamefont{{Yahyaoui}}}, \bibinfo{journal}{\aap{}}
  \textbf{\bibinfo{volume}{521}}, \bibinfo{pages}{A47} (\bibinfo{year}{2010}).

\bibitem[{\citenamefont{{James}}(2012)}]{james2012}
\bibinfo{author}{\bibfnamefont{C.~W.} \bibnamefont{{James}}}, in
  \emph{\bibinfo{booktitle}{Proc.\ ARENA 2010}} (\bibinfo{year}{2012}), vol.
  \bibinfo{volume}{662} of \emph{\bibinfo{series}{\nima{}}},
  p.~\bibinfo{pages}{12}.

\bibitem[{\citenamefont{{Gayley} et~al.}(2009)\citenamefont{{Gayley}, {Mutel},
  and {Jaeger}}}]{gayley2009}
\bibinfo{author}{\bibfnamefont{K.~G.} \bibnamefont{{Gayley}}},
  \bibinfo{author}{\bibfnamefont{R.~L.} \bibnamefont{{Mutel}}},
  \bibnamefont{and} \bibinfo{author}{\bibfnamefont{T.~R.}
  \bibnamefont{{Jaeger}}}, \bibinfo{journal}{\apj{}}
  \textbf{\bibinfo{volume}{706}}, \bibinfo{pages}{1556} (\bibinfo{year}{2009}).

\bibitem[{\citenamefont{{Scholten} et~al.}(2006)\citenamefont{{Scholten},
  {Bacelar}, {Braun}, {de Bruyn}, {Falcke}, {Stappers}, and
  {Strom}}}]{scholten2006}
\bibinfo{author}{\bibfnamefont{O.}~\bibnamefont{{Scholten}}},
  \bibinfo{author}{\bibfnamefont{J.}~\bibnamefont{{Bacelar}}},
  \bibinfo{author}{\bibfnamefont{R.}~\bibnamefont{{Braun}}},
  \bibinfo{author}{\bibfnamefont{A.~G.} \bibnamefont{{de Bruyn}}},
  \bibinfo{author}{\bibfnamefont{H.}~\bibnamefont{{Falcke}}},
  \bibinfo{author}{\bibfnamefont{B.}~\bibnamefont{{Stappers}}},
  \bibnamefont{and} \bibinfo{author}{\bibfnamefont{R.~G.}
  \bibnamefont{{Strom}}}, \bibinfo{journal}{\app{}}
  \textbf{\bibinfo{volume}{26}}, \bibinfo{pages}{219} (\bibinfo{year}{2006}).

\bibitem[{\citenamefont{{Jaeger} et~al.}(2010)\citenamefont{{Jaeger}, {Mutel},
  and {Gayley}}}]{jaeger2010}
\bibinfo{author}{\bibfnamefont{T.~R.} \bibnamefont{{Jaeger}}},
  \bibinfo{author}{\bibfnamefont{R.~L.} \bibnamefont{{Mutel}}},
  \bibnamefont{and} \bibinfo{author}{\bibfnamefont{K.~G.}
  \bibnamefont{{Gayley}}}, \bibinfo{journal}{\app{}}
  \textbf{\bibinfo{volume}{34}}, \bibinfo{pages}{293} (\bibinfo{year}{2010}).

\bibitem[{\citenamefont{{Gorham} et~al.}(2010)\citenamefont{{Gorham},
  {Allison}, {Baughman}, {Beatty}, {Belov}, {Besson}, {Bevan}, {Binns}, {Chen},
  {Chen} et~al.}}]{gorham2010}
\bibinfo{author}{\bibfnamefont{P.~W.} \bibnamefont{{Gorham}}},
  \bibinfo{author}{\bibfnamefont{P.}~\bibnamefont{{Allison}}},
  \bibinfo{author}{\bibfnamefont{B.~M.} \bibnamefont{{Baughman}}},
  \bibinfo{author}{\bibfnamefont{J.~J.} \bibnamefont{{Beatty}}},
  \bibinfo{author}{\bibfnamefont{K.}~\bibnamefont{{Belov}}},
  \bibinfo{author}{\bibfnamefont{D.~Z.} \bibnamefont{{Besson}}},
  \bibinfo{author}{\bibfnamefont{S.}~\bibnamefont{{Bevan}}},
  \bibinfo{author}{\bibfnamefont{W.~R.} \bibnamefont{{Binns}}},
  \bibinfo{author}{\bibfnamefont{C.}~\bibnamefont{{Chen}}},
  \bibinfo{author}{\bibfnamefont{P.}~\bibnamefont{{Chen}}},
  \bibnamefont{et~al.} (\bibinfo{collaboration}{{ANITA Collaboration}}),
  \bibinfo{journal}{\prd{}} \textbf{\bibinfo{volume}{82}},
  \bibinfo{pages}{022004} (\bibinfo{year}{2010}).

\bibitem[{\citenamefont{{Kravchenko}}(2012)}]{kravchenko2012}
\bibinfo{author}{\bibfnamefont{I.}~\bibnamefont{{Kravchenko}}}
  (\bibinfo{collaboration}{{RICE Collaboration}}), in
  \emph{\bibinfo{booktitle}{Proc.\ ARENA 2010}} (\bibinfo{year}{2012}), vol.
  \bibinfo{volume}{662} of \emph{\bibinfo{series}{\nima{}}},
  p.~\bibinfo{pages}{42}.

\bibitem[{\citenamefont{{Aartsen} et~al.}(2013)\citenamefont{{Aartsen},
  {Abbasi}, {Ackermann}, {Adams}, {Aguilar}, {Ahlers}, {Altmann}, {Arguelles},
  {Auffenberg}, {Bai} et~al.}}]{aartsen2013c}
\bibinfo{author}{\bibfnamefont{M.~G.} \bibnamefont{{Aartsen}}},
  \bibinfo{author}{\bibfnamefont{R.}~\bibnamefont{{Abbasi}}},
  \bibinfo{author}{\bibfnamefont{M.}~\bibnamefont{{Ackermann}}},
  \bibinfo{author}{\bibfnamefont{J.}~\bibnamefont{{Adams}}},
  \bibinfo{author}{\bibfnamefont{J.~A.} \bibnamefont{{Aguilar}}},
  \bibinfo{author}{\bibfnamefont{M.}~\bibnamefont{{Ahlers}}},
  \bibinfo{author}{\bibfnamefont{D.}~\bibnamefont{{Altmann}}},
  \bibinfo{author}{\bibfnamefont{C.}~\bibnamefont{{Arguelles}}},
  \bibinfo{author}{\bibfnamefont{J.}~\bibnamefont{{Auffenberg}}},
  \bibinfo{author}{\bibfnamefont{X.}~\bibnamefont{{Bai}}}, \bibnamefont{et~al.}
  (\bibinfo{collaboration}{IceCube Collaboration}), \bibinfo{journal}{\prd{}}
  \textbf{\bibinfo{volume}{88}}, \bibinfo{pages}{112008}
  (\bibinfo{year}{2013}).

\bibitem[{\citenamefont{{Abreu} et~al.}(2012)\citenamefont{{Abreu}, {Aglietta},
  {Ahlers}, {Ahn}, {Albuquerque}, {Allard}, {Allekotte}, {Allen}, {Allison},
  {Almela} et~al.}}]{abreu2012}
\bibinfo{author}{\bibfnamefont{P.}~\bibnamefont{{Abreu}}},
  \bibinfo{author}{\bibfnamefont{M.}~\bibnamefont{{Aglietta}}},
  \bibinfo{author}{\bibfnamefont{M.}~\bibnamefont{{Ahlers}}},
  \bibinfo{author}{\bibfnamefont{E.~J.} \bibnamefont{{Ahn}}},
  \bibinfo{author}{\bibfnamefont{I.~F.~M.} \bibnamefont{{Albuquerque}}},
  \bibinfo{author}{\bibfnamefont{D.}~\bibnamefont{{Allard}}},
  \bibinfo{author}{\bibfnamefont{I.}~\bibnamefont{{Allekotte}}},
  \bibinfo{author}{\bibfnamefont{J.}~\bibnamefont{{Allen}}},
  \bibinfo{author}{\bibfnamefont{P.}~\bibnamefont{{Allison}}},
  \bibinfo{author}{\bibfnamefont{A.}~\bibnamefont{{Almela}}},
  \bibnamefont{et~al.} (\bibinfo{collaboration}{{Pierre Auger Collaboration}}),
  \bibinfo{journal}{\apjl{}} \textbf{\bibinfo{volume}{755}}, \bibinfo{eid}{L4}
  (\bibinfo{year}{2012}).

\bibitem[{\citenamefont{{Abreu} et~al.}(2011)\citenamefont{{Abreu}, {Aglietta},
  {Ahn}, {Albuquerque}, {Allard}, {Allekotte}, {Allen}, {Allison}, {Alvarez
  Castillo}, {Alvarez-Mu{\~n}iz} et~al.}}]{abreu2011b}
\bibinfo{author}{\bibfnamefont{P.}~\bibnamefont{{Abreu}}},
  \bibinfo{author}{\bibfnamefont{M.}~\bibnamefont{{Aglietta}}},
  \bibinfo{author}{\bibfnamefont{E.~J.} \bibnamefont{{Ahn}}},
  \bibinfo{author}{\bibfnamefont{I.~F.~M.} \bibnamefont{{Albuquerque}}},
  \bibinfo{author}{\bibfnamefont{D.}~\bibnamefont{{Allard}}},
  \bibinfo{author}{\bibfnamefont{I.}~\bibnamefont{{Allekotte}}},
  \bibinfo{author}{\bibfnamefont{J.}~\bibnamefont{{Allen}}},
  \bibinfo{author}{\bibfnamefont{P.}~\bibnamefont{{Allison}}},
  \bibinfo{author}{\bibfnamefont{J.}~\bibnamefont{{Alvarez Castillo}}},
  \bibinfo{author}{\bibfnamefont{J.}~\bibnamefont{{Alvarez-Mu{\~n}iz}}},
  \bibnamefont{et~al.} (\bibinfo{collaboration}{Pierre Auger Collaboration}),
  \bibinfo{journal}{\prd{}} \textbf{\bibinfo{volume}{84}},
  \bibinfo{pages}{122005} (\bibinfo{year}{2011}).

\bibitem[{\citenamefont{{Allard} et~al.}(2006)\citenamefont{{Allard}, {Ave},
  {Busca}, {Malkan}, {Olinto}, {Parizot}, {Stecker}, and
  {Yamamoto}}}]{allard2006}
\bibinfo{author}{\bibfnamefont{D.}~\bibnamefont{{Allard}}},
  \bibinfo{author}{\bibfnamefont{M.}~\bibnamefont{{Ave}}},
  \bibinfo{author}{\bibfnamefont{N.}~\bibnamefont{{Busca}}},
  \bibinfo{author}{\bibfnamefont{M.~A.} \bibnamefont{{Malkan}}},
  \bibinfo{author}{\bibfnamefont{A.~V.} \bibnamefont{{Olinto}}},
  \bibinfo{author}{\bibfnamefont{E.}~\bibnamefont{{Parizot}}},
  \bibinfo{author}{\bibfnamefont{F.~W.} \bibnamefont{{Stecker}}},
  \bibnamefont{and}
  \bibinfo{author}{\bibfnamefont{T.}~\bibnamefont{{Yamamoto}}},
  \bibinfo{journal}{\jcap{}} \textbf{\bibinfo{volume}{9}}, \bibinfo{pages}{5}
  (\bibinfo{year}{2006}).

\bibitem[{\citenamefont{{Abraham}
  et~al.}(2010{\natexlab{b}})\citenamefont{{Abraham}, {Abreu}, {Aglietta},
  {Ahn}, {Allard}, {Allekotte}, {Allen}, {Alvarez-Mu{\~n}iz}, {Ambrosio},
  {Anchordoqui} et~al.}}]{abraham2010b}
\bibinfo{author}{\bibfnamefont{J.}~\bibnamefont{{Abraham}}},
  \bibinfo{author}{\bibfnamefont{P.}~\bibnamefont{{Abreu}}},
  \bibinfo{author}{\bibfnamefont{M.}~\bibnamefont{{Aglietta}}},
  \bibinfo{author}{\bibfnamefont{E.~J.} \bibnamefont{{Ahn}}},
  \bibinfo{author}{\bibfnamefont{D.}~\bibnamefont{{Allard}}},
  \bibinfo{author}{\bibfnamefont{I.}~\bibnamefont{{Allekotte}}},
  \bibinfo{author}{\bibfnamefont{J.}~\bibnamefont{{Allen}}},
  \bibinfo{author}{\bibfnamefont{J.}~\bibnamefont{{Alvarez-Mu{\~n}iz}}},
  \bibinfo{author}{\bibfnamefont{M.}~\bibnamefont{{Ambrosio}}},
  \bibinfo{author}{\bibfnamefont{L.}~\bibnamefont{{Anchordoqui}}},
  \bibnamefont{et~al.} (\bibinfo{collaboration}{Pierre Auger Collaboration}),
  \bibinfo{journal}{\prl{}} \textbf{\bibinfo{volume}{104}},
  \bibinfo{pages}{091101} (\bibinfo{year}{2010}{\natexlab{b}}).

\bibitem[{\citenamefont{{Shepard} et~al.}(1995)\citenamefont{{Shepard},
  {Brackett}, and {Arvidson}}}]{shepard1995}
\bibinfo{author}{\bibfnamefont{M.~K.} \bibnamefont{{Shepard}}},
  \bibinfo{author}{\bibfnamefont{R.~A.} \bibnamefont{{Brackett}}},
  \bibnamefont{and} \bibinfo{author}{\bibfnamefont{R.~E.}
  \bibnamefont{{Arvidson}}}, \bibinfo{journal}{\jgr{}}
  \textbf{\bibinfo{volume}{100}}, \bibinfo{pages}{11709}
  (\bibinfo{year}{1995}).

\bibitem[{\citenamefont{{Fanaroff} and {Riley}}(1974)}]{fanaroff1974}
\bibinfo{author}{\bibfnamefont{B.~L.} \bibnamefont{{Fanaroff}}}
  \bibnamefont{and} \bibinfo{author}{\bibfnamefont{J.~M.}
  \bibnamefont{{Riley}}}, \bibinfo{journal}{\mnras{}}
  \textbf{\bibinfo{volume}{167}}, \bibinfo{pages}{31P} (\bibinfo{year}{1974}).

\bibitem[{\citenamefont{{Harris} et~al.}(2010)\citenamefont{{Harris},
  {Rejkuba}, and {Harris}}}]{harris2010}
\bibinfo{author}{\bibfnamefont{G.~L.~H.} \bibnamefont{{Harris}}},
  \bibinfo{author}{\bibfnamefont{M.}~\bibnamefont{{Rejkuba}}},
  \bibnamefont{and} \bibinfo{author}{\bibfnamefont{W.~E.}
  \bibnamefont{{Harris}}}, \bibinfo{journal}{\pasa{}}
  \textbf{\bibinfo{volume}{27}}, \bibinfo{pages}{457} (\bibinfo{year}{2010}).

\bibitem[{\citenamefont{{Allard} and {Protheroe}}(2009)}]{allard2009}
\bibinfo{author}{\bibfnamefont{D.}~\bibnamefont{{Allard}}} \bibnamefont{and}
  \bibinfo{author}{\bibfnamefont{R.~J.} \bibnamefont{{Protheroe}}},
  \bibinfo{journal}{\aap{}} \textbf{\bibinfo{volume}{502}},
  \bibinfo{pages}{803} (\bibinfo{year}{2009}).

\bibitem[{\citenamefont{{Aartsen} et~al.}(2014)\citenamefont{{Aartsen},
  {Ackermann}, {Adams}, {Aguilar}, {Ahlers}, {Ahrens}, {Altmann}, {Anderson},
  {Arguelles}, {Arlen} et~al.}}]{aartsen2014}
\bibinfo{author}{\bibfnamefont{M.~G.} \bibnamefont{{Aartsen}}},
  \bibinfo{author}{\bibfnamefont{M.}~\bibnamefont{{Ackermann}}},
  \bibinfo{author}{\bibfnamefont{J.}~\bibnamefont{{Adams}}},
  \bibinfo{author}{\bibfnamefont{J.~A.} \bibnamefont{{Aguilar}}},
  \bibinfo{author}{\bibfnamefont{M.}~\bibnamefont{{Ahlers}}},
  \bibinfo{author}{\bibfnamefont{M.}~\bibnamefont{{Ahrens}}},
  \bibinfo{author}{\bibfnamefont{D.}~\bibnamefont{{Altmann}}},
  \bibinfo{author}{\bibfnamefont{T.}~\bibnamefont{{Anderson}}},
  \bibinfo{author}{\bibfnamefont{C.}~\bibnamefont{{Arguelles}}},
  \bibinfo{author}{\bibfnamefont{T.~C.} \bibnamefont{{Arlen}}},
  \bibnamefont{et~al.} (\bibinfo{collaboration}{IceCube Collaboration}),
  \bibinfo{journal}{\apj{}} \textbf{\bibinfo{volume}{796}},
  \bibinfo{pages}{109} (\bibinfo{year}{2014}).

\bibitem[{\citenamefont{{Besson}}(2009)}]{besson2009}
\bibinfo{author}{\bibfnamefont{D.}~\bibnamefont{{Besson}}},
  \bibinfo{howpublished}{\privcom{}} (\bibinfo{year}{2009}).

\bibitem[{\citenamefont{{Cuoco} and {Hannestad}}(2008)}]{cuoco2008}
\bibinfo{author}{\bibfnamefont{A.}~\bibnamefont{{Cuoco}}} \bibnamefont{and}
  \bibinfo{author}{\bibfnamefont{S.}~\bibnamefont{{Hannestad}}},
  \bibinfo{journal}{\prd{}} \textbf{\bibinfo{volume}{78}},
  \bibinfo{pages}{023007} (\bibinfo{year}{2008}).

\bibitem[{\citenamefont{{Kachelrie{\ss}}
  et~al.}(2009{\natexlab{a}})\citenamefont{{Kachelrie{\ss}}, {Ostapchenko}, and
  {Tom{\`a}s}}}]{kachelriess2009}
\bibinfo{author}{\bibfnamefont{M.}~\bibnamefont{{Kachelrie{\ss}}}},
  \bibinfo{author}{\bibfnamefont{S.}~\bibnamefont{{Ostapchenko}}},
  \bibnamefont{and}
  \bibinfo{author}{\bibfnamefont{R.}~\bibnamefont{{Tom{\`a}s}}},
  \bibinfo{journal}{\njp{}} \textbf{\bibinfo{volume}{11}},
  \bibinfo{pages}{065017} (\bibinfo{year}{2009}{\natexlab{a}}).

\bibitem[{\citenamefont{{Kachelrie{\ss}}
  et~al.}(2009{\natexlab{b}})\citenamefont{{Kachelrie{\ss}}, {Ostapchenko}, and
  {Tom{\`a}s}}}]{kachelriess2009b}
\bibinfo{author}{\bibfnamefont{M.}~\bibnamefont{{Kachelrie{\ss}}}},
  \bibinfo{author}{\bibfnamefont{S.}~\bibnamefont{{Ostapchenko}}},
  \bibnamefont{and}
  \bibinfo{author}{\bibfnamefont{R.}~\bibnamefont{{Tom{\`a}s}}},
  \bibinfo{journal}{\ijmpd{}} \textbf{\bibinfo{volume}{18}},
  \bibinfo{pages}{1591} (\bibinfo{year}{2009}{\natexlab{b}}).

\bibitem[{\citenamefont{{Hillas}}(1984)}]{hillas1984}
\bibinfo{author}{\bibfnamefont{A.~M.} \bibnamefont{{Hillas}}},
  \bibinfo{journal}{\araa{}} \textbf{\bibinfo{volume}{22}},
  \bibinfo{pages}{425} (\bibinfo{year}{1984}).

\bibitem[{\citenamefont{{Koers} and {Tinyakov}}(2008)}]{koers2008}
\bibinfo{author}{\bibfnamefont{H.~B.~J.} \bibnamefont{{Koers}}}
  \bibnamefont{and}
  \bibinfo{author}{\bibfnamefont{P.}~\bibnamefont{{Tinyakov}}},
  \bibinfo{journal}{\prd{}} \textbf{\bibinfo{volume}{78}},
  \bibinfo{pages}{083009} (\bibinfo{year}{2008}).

\bibitem[{\citenamefont{{Waxman} and {Bahcall}}(1997)}]{waxman1997}
\bibinfo{author}{\bibfnamefont{E.}~\bibnamefont{{Waxman}}} \bibnamefont{and}
  \bibinfo{author}{\bibfnamefont{J.}~\bibnamefont{{Bahcall}}},
  \bibinfo{journal}{\prl{}} \textbf{\bibinfo{volume}{78}},
  \bibinfo{pages}{2292} (\bibinfo{year}{1997}).

\bibitem[{\citenamefont{{Abbasi} et~al.}(2012)\citenamefont{{Abbasi}, {Abdou},
  {Abu-Zayyad}, {Ackermann}, {Adams}, {Aguilar}, {Ahlers}, {Altmann}, {Andeen},
  {Auffenberg} et~al.}}]{abbasi2012}
\bibinfo{author}{\bibfnamefont{R.}~\bibnamefont{{Abbasi}}},
  \bibinfo{author}{\bibfnamefont{Y.}~\bibnamefont{{Abdou}}},
  \bibinfo{author}{\bibfnamefont{T.}~\bibnamefont{{Abu-Zayyad}}},
  \bibinfo{author}{\bibfnamefont{M.}~\bibnamefont{{Ackermann}}},
  \bibinfo{author}{\bibfnamefont{J.}~\bibnamefont{{Adams}}},
  \bibinfo{author}{\bibfnamefont{J.~A.} \bibnamefont{{Aguilar}}},
  \bibinfo{author}{\bibfnamefont{M.}~\bibnamefont{{Ahlers}}},
  \bibinfo{author}{\bibfnamefont{D.}~\bibnamefont{{Altmann}}},
  \bibinfo{author}{\bibfnamefont{K.}~\bibnamefont{{Andeen}}},
  \bibinfo{author}{\bibfnamefont{J.}~\bibnamefont{{Auffenberg}}},
  \bibnamefont{et~al.} (\bibinfo{collaboration}{IceCube Collaboration}),
  \bibinfo{journal}{\nat{}} \textbf{\bibinfo{volume}{484}},
  \bibinfo{pages}{351} (\bibinfo{year}{2012}).

\bibitem[{\citenamefont{{H{\"u}mmer} et~al.}(2012)\citenamefont{{H{\"u}mmer},
  {Baerwald}, and {Winter}}}]{hummer2012}
\bibinfo{author}{\bibfnamefont{S.}~\bibnamefont{{H{\"u}mmer}}},
  \bibinfo{author}{\bibfnamefont{P.}~\bibnamefont{{Baerwald}}},
  \bibnamefont{and} \bibinfo{author}{\bibfnamefont{W.}~\bibnamefont{{Winter}}},
  \bibinfo{journal}{\prl{}} \textbf{\bibinfo{volume}{108}},
  \bibinfo{eid}{231101} (\bibinfo{year}{2012}).

\bibitem[{\citenamefont{{Barthelmy} et~al.}(2000)\citenamefont{{Barthelmy},
  {Cline}, {Butterworth}, {Kippen}, {Briggs}, {Connaughton}, and
  {Pendleton}}}]{barthelmy2000}
\bibinfo{author}{\bibfnamefont{S.~D.} \bibnamefont{{Barthelmy}}},
  \bibinfo{author}{\bibfnamefont{T.~L.} \bibnamefont{{Cline}}},
  \bibinfo{author}{\bibfnamefont{P.}~\bibnamefont{{Butterworth}}},
  \bibinfo{author}{\bibfnamefont{R.~M.} \bibnamefont{{Kippen}}},
  \bibinfo{author}{\bibfnamefont{M.~S.} \bibnamefont{{Briggs}}},
  \bibinfo{author}{\bibfnamefont{V.}~\bibnamefont{{Connaughton}}},
  \bibnamefont{and} \bibinfo{author}{\bibfnamefont{G.~N.}
  \bibnamefont{{Pendleton}}}, in \emph{\bibinfo{booktitle}{Proc.\ Huntsville
  GRB Symp.\ 1999}} (\bibinfo{year}{2000}), vol. \bibinfo{volume}{526} of
  \emph{\bibinfo{series}{\aipcs{}}}, pp. \bibinfo{pages}{731--735},
  \urlprefix\url{http://gcn.gsfc.nasa.gov/}.

\end{thebibliography}

\end{document}